\documentclass[aps,twocolumn,amsmath,amssymb,superscriptaddress]{revtex4-2}
\pdfoutput=1
\usepackage[colorlinks=true,linkcolor=blue,citecolor=blue,urlcolor=blue]{hyperref}

\usepackage{amsmath}
\usepackage{graphicx}
\usepackage{graphics}
\usepackage{bm}
\usepackage[usenames]{color}
\usepackage{colordvi}
\usepackage{color}
\usepackage{units}
\usepackage{bbm}
\usepackage{booktabs}
\usepackage{array}
\usepackage{placeins}
\usepackage{epsfig}
\usepackage{changes}
\usepackage{braket}
\usepackage{tabularx}
\newcolumntype{L}[1]{>{\raggedright\arraybackslash}p{#1}} % linksbündig mit Breitenangabe
\newcolumntype{C}[1]{>{\centering\arraybackslash}p{#1}} % zentriert mit Breitenangabe
\newcolumntype{R}[1]{>{\raggedleft\arraybackslash}p{#1}} % rechtsbündig mit Breitenangabe
\usepackage[colorlinks=true,linkcolor=blue,citecolor=blue,urlcolor=blue]{hyperref}
\newcommand{\be}{\begin{equation}}
\newcommand{\ee}{\end{equation}}
\newcommand{\beqn}{\begin{eqnarray}}
\newcommand{\eeqn}{\end{eqnarray}}

\definecolor{mymagenta}{rgb}{1.0,0.0,1.0}
\definecolor{mycyan}{rgb}{0.0,1.0,1.0}
\definecolor{myyellow}{rgb}{1.0,1.0,0.0}
\definecolor{myorange}{rgb}{1.0,0.27,0.0}

\definecolor{dark-gray}{HTML}{a0a0a0}
\definecolor{dark-red}{HTML}{8b0000}
\definecolor{dark-green}{HTML}{006400}
\definecolor{dark-blue}{HTML}{00008b}
\definecolor{gold}{rgb}{1.0,0.84,0.0}
\definecolor{dark-turquoise}{HTML}{00ced1}

\bibstyle{apsrev.bib}

\begin{document}

\title{Random transverse and longitudinal field Ising chains}
\author{Tam\'as Pet\H {o}}
\email{petotamas0@gmail.com}
\affiliation{Institute of Theoretical Physics, Szeged University, H-6720 Szeged, Hungary}
\author{Ferenc Igl{\'o}i}
\email{igloi.ferenc@wigner.hu}
\affiliation{HUN-REN Wigner Research Centre for Physics, Institute for Solid State Physics and Optics, H-1525 Budapest, P.O. Box 49, Hungary}
\affiliation{Institute of Theoretical Physics, Szeged University, H-6720 Szeged, Hungary}
\author{Istv\'an A. Kov\'acs}
\email{istvan.kovacs@northwestern.edu}
\affiliation{Department of Physics and Astronomy, Northwestern University, Evanston, IL 60208, USA}
\affiliation{Northwestern Institute on Complex Systems, Northwestern University, Evanston, IL 60208, USA}
\affiliation{Department of Engineering Sciences and Applied Mathematics, Northwestern University, Evanston, IL 60208, USA}

\date{\today}

\begin{abstract}
 Motivated by experimental results on compounds like ${\rm LiHo}_x{\rm Y}_{1-x}{\rm F}_4$, we consider an Ising chain with random bonds in the simultaneous presence of random transverse and longitudinal fields. We study the low-energy properties of the model at zero temperature by the strong disorder renormalization group method.  In the absence of random longitudinal fields, the model showcases a trivial quantum-ordered and quantum-disordered fixed-point and a non-trivial infinite disorder critical point. In the absence of random transverse fields, the behavior is  dictated by the classical random-field Ising fixed-point.  
In the simultaneous presence of both a longitudinal and transverse random field, the RG trajectories are attracted to a set of disordered fixed-points, in which the disorder is either due to random quantum fluctuations, or due to classical random-field effects.  Between the two regimes there is a smooth cross-over, which becomes sharp at the infinite disorder fixed-point. This local separatrix defines the relevant scaling direction, where the correlation-length is shown to diverge with an exponent $\nu_h \approx 1$.
\end{abstract}


\maketitle

\section{Introduction}
\label{sec:intr}
Quantum phase transitions formally take place at zero temperature by varying a control-parameter, such as the strength of a transverse magnetic field\cite{sachdev_2011}. In a $d$-dimensional quantum system the phase transition is often related to a classical one in $(d+1)$-dimensions, such as in the case of the transverse-field Ising chain and the two-dimensional classical Ising model. Quantum phase transitions, however, can be also different from the existing classical ones, like in the case of the so called deconfined criticality\cite{senthil2023deconfinedquantumcriticalpoints}. The effects of a quantum phase transition are manifested also at low, but finite temperatures, where several physical observables can show singular characteristics. 

Quenched disorder is an inevitable feature of real materials and it can have a profound effect on the properties of the quantum phase transition. A frequently occurring scenario is given by random models in which the phase transition is controlled by a so called infinite disorder fixed-point (IDFP), the properties of which are completely dominated by disorder fluctuations\cite{FISHER1999222}. The prototype of such systems is the random transverse-field Ising chain, the critical properties of which has been calculated by Daniel Fisher\cite{PhysRevLett.69.534,PhysRevB.51.6411} by the use of a so called strong disorder renormalization group (SDRG) method\cite{IGLOI2005277,Igloi2018}, extending on an original idea by Ma, Dasgupta and Hu\cite{PhysRevLett.43.1434,PhysRevB.22.1305}. The SDRG technique %the decimated degrees of freedom are selected 
operates in an (excitation) energy basis: in each step, the parameter associated with the highest local excitation energy is decimated. The local decimation then leads to the creation of new small parameters that are calculated perturbatively between the remaining sites. As the renormalization is iterated the energy-scale goes to zero and the fixed-point of the transformation will control the properties of the phase transition. As shown by Fisher, at the IDFP the perturbative steps become asymptotically exact and it is expected that the fixed-point describes the correct critical behaviour of the system\cite{PhysRevLett.69.534,PhysRevB.51.6411}.

In one-dimensional models -- where the (chain) topology of the system remains invariant under renormalization -- the RG-flow equations can be written in a set of integro-differential equations and can be solved analytically. Examples are the random transverse-field Ising chain\cite{PhysRevLett.69.534,PhysRevB.51.6411}, the random $XX$ and $XXZ$ chains\cite{PhysRevB.50.3799} and several other systems\cite{PhysRevLett.76.3001,PhysRevLett.87.277201} for reviews see\cite{IGLOI2005277,Igloi2018}. In these systems, the SDRG results are generally confronted with detailed numerical calculations\cite{PhysRevB.53.8486,PhysRevB.57.11404,PhysRevB.58.9131,PhysRevB.61.11552}, and a good agreement is obtained. In higher dimensions, the topology of the system changes during the renormalization process and the calculations need to be performed numerically\cite{PhysRevB.61.1160,10.1143/PTPS.138.479,Karevski2001,PhysRevLett.99.147202,PhysRevB.77.140402,PhysRevB.80.214416}. Several efficient numerical algorithms have been developed\cite{PhysRevB.82.054437,PhysRevB.83.174207,Kov_cs_2011}, so that systems with considerably large linear extent could be accurately renormalized. The obtained results indicate that the transverse-field Ising model (with nearest neighbour couplings) has an IDFP in any spatial dimensions\cite{PhysRevB.83.174207,Kov_cs_2011}, including various network topologies\cite{Juhasz_2013} and this fixed-point is likely to control the critical behaviour of any other models having a discrete order-parameter variable\cite{PhysRevLett.76.3001,PhysRevLett.87.277201,PhysRevB.103.174207,PhysRevE.108.014124}. This scenario changes for models with long-range forces, where the critical fixed-point is conventional random\cite{Juh_sz_2014,PhysRevB.93.184203}, or for models with three-spin product interactions\cite{e26080709}.

The random transverse-field Ising model has several experimental realizations, including order-disorder ferroelectrics
($\rm{K}(\rm{H}_x\rm{D}_{1-x})_2\rm{PO}_4$)\cite{doi:10.1080/00018736900101297,PhysRevLett.20.1105,doi:10.1143/JPSJ.24.497,PhysRevLett.27.103}, mixed hydrogen bonded ferroelectrics ($\rm{Rd}_{1-x}\rm{(NH_4)}_x\rm{H}_2\rm{PO}_4$)\cite{Pirc1985}, quasi-1D Ising systems ($\rm{CoNb}_2\rm{O}_6$)\cite{doi:10.1126/science.1180085} and dipolar magnets ${\rm LiHo}_x{\rm Y}_{1-x}{\rm F}_4$. For a more extensive list, see\cite{RBStinchcombe_1973}.
Among these, the most data is available for the   ${\rm LiHo}_x{\rm Y}_{1-x}{\rm F}_4$ compound, in which a fraction of $(1-x)$ of the magnetic Ho atoms is replaced by nonmagnetic Y atoms\cite{PhysRevB.42.4631,PhysRevLett.67.2076,PhysRevLett.71.1919,science.284.5415.779,dutta_aeppli_chakrabarti_divakaran_rosenbaum_sen_2015}.
If this system is placed into a magnetic field which is transverse to the Ising axis it acts as an effective transverse field. The low-energy properties of this system are well described by a random transverse-field Ising model (with long-range interactions), but in this compound the transverse field also induces a random longitudinal field via the off-diagonal terms of the dipolar interaction\cite{PhysRevLett.97.137204,PhysRevLett.97.237203,PhysRevB.77.020401,Schechter_2009}. We also mention recent experimental progress in the superconductor-metal transition and the accompanying quantum Griffiths singularity\cite{Xing2023,Yadav2024,Wang_2024,10.1093/nsr/nwae220,doi:10.1126/sciadv.adp1402,PhysRevLett.133.226001}. 

Motivated by the  ${\rm LiHo}_x{\rm Y}_{1-x}{\rm F}_4$ compound, it is natural to consider an Ising model that contains random transverse and random longitudinal fields at the same time. In this paper, we study this system with nearest neighbour interactions in one dimension, given by the Hamiltonian:
\begin{align}
\begin{split}
\hat{H}&=-\sum_{i=1}^L J_i \sigma_{i}^{z} \sigma_{i+1}^{z}\\
&-\sum_{i=1}^L \Gamma_i \sigma^x_{j} -\sum_{i=1}^L {\textcolor{black}{c_i}}h_i \sigma^z_{i}\;,
\label{Hamilton}
\end{split}
\end{align}
Here the $\sigma_{i}^{x,z}$ are Pauli matrices at site $i$ and we use periodic boundary conditions: $\sigma_{L+1}^{z}\equiv \sigma_{1}^{z}$. The nearest neighbour couplings are ferromagnetic, $J_i>0$ and random, the transverse fields, $\Gamma_i>0$ are random, too. For the longitudinal field, we assume that it acts only on a fraction of sites, $\zeta \le 1$, {\textcolor{black}{thus}
\textcolor{black}{
\be
c_i=
   \begin{cases}
     1 & \hspace*{0.65cm}\text{with probability }~ \zeta\,,\\
     0 & \hspace*{0.65cm}\text{with probability}~ (1-\zeta)\;.
    \end{cases}
\label{rho_i}
\ee
}
{\textcolor{black}{The distribution of the longitudinal fields is symmetric: $p(h)=p(-h)$}. 
Throughout the paper we used the following box-like distributions in the calculations:
\begin{align}
  \begin{split}
    \pi(J) &=
    \begin{cases}
      1 & \hspace*{0.65cm}\text{for } 0<J\le 1\,,\\
      0 & \hspace*{0.65cm}\text{otherwise.}
    \end{cases} \\
    \pi_2(\Gamma) &= 
    \begin{cases}
      1/\Gamma_0& \text{for } 0<\Gamma \le \Gamma_0\,,\\
      0 & \text{otherwise.} 
    \end{cases}\\
    p(h) &= 
    \begin{cases}
      1/h_0& \text{for } -h_0/2 \le h \le h_0/2\,,\\
      0 & \text{otherwise.} 
    \end{cases}\\
  \end{split} 
  \label{eq:J_distrib} 
\end{align}

Note that alternative variants of the model in Eq.(\ref{Hamilton}) have been also of interest. Setting $J_i=J$ and $\Gamma_i=\Gamma$ while the longitudinal field is $h_i=h(-1)^i$ is equivalent to the antiferromagnetic Ising model in transverse and longitudinal fields. This model has been studied theoretically in Refs.\cite{PhysRevE.63.016112,PhysRevB.68.214406,PhysRevE.99.012122,PhysRevB.103.174404} and experimentally in Ref.\cite{Simon2011}. For random couplings and random transverse fields, but with non-random staggered longitudinal fields it is studied in Refs.\cite{PhysRevB.96.064427,PhysRevB.101.024203} and a reentrant random quantum Ising antiferromagnetic phase is observed.

We studied the cooperative properties of the model in Eq.(\ref{Hamilton}) by the SDRG method and obtained a schematic phase-diagram which is shown in Fig.\ref{fig_1}. 
As seen in the figure, the system has an ordered phase at $h_0=0$ and for $\Gamma_0<\Gamma_0^c$, which is controlled by a trivial fixed-point at $\Gamma_0=0$ and indicated by a black circle. For other values of the parameters the system has no long-range order \textcolor{black}{and the RG trajectories are attracted by a set of disordered fixed points, in which the couplings are negligible and the disorder is either due to random quantum fluctuations (illustrated by blue trajectories), or due to classical random field effects (illustrated by green trajectories).  Between the two regimes there is a smooth cross-over region, the borders of which are illustrated by dashed red lines, where the two random fields play practically equivalent role.  At $h_0=0$ the quantum ordered phase and the quantum disordered phase (having a trivial fixed-point at $\Gamma_0 \to \infty$ and indicated by a blue circle) is separated by an IDFP at $\Gamma_0=\Gamma_0^c$. Note that for the distribution in Eq.(\ref{eq:J_distrib}) it is known exactly\cite{PhysRevLett.69.534,PhysRevB.51.6411} that  $\Gamma_0^c=1$. We shall show that close to the IDFP for $h_0 \to 0$ the cross-over is sharp and this local separatrix defines the relevant scaling direction. In the classical limit, $\Gamma_0=0$ and for $h_0>0$ the flows are controlled by the fixed point of the random-field Ising model, which is located at $h_0 \to \infty$ and indicated by a green circle.}
%%%%%%%%%%%%%%%%%%%%%%%%% Fig 1b %%%%%%%%%%
\begin{figure}[h!]
\includegraphics[width=1. \columnwidth]{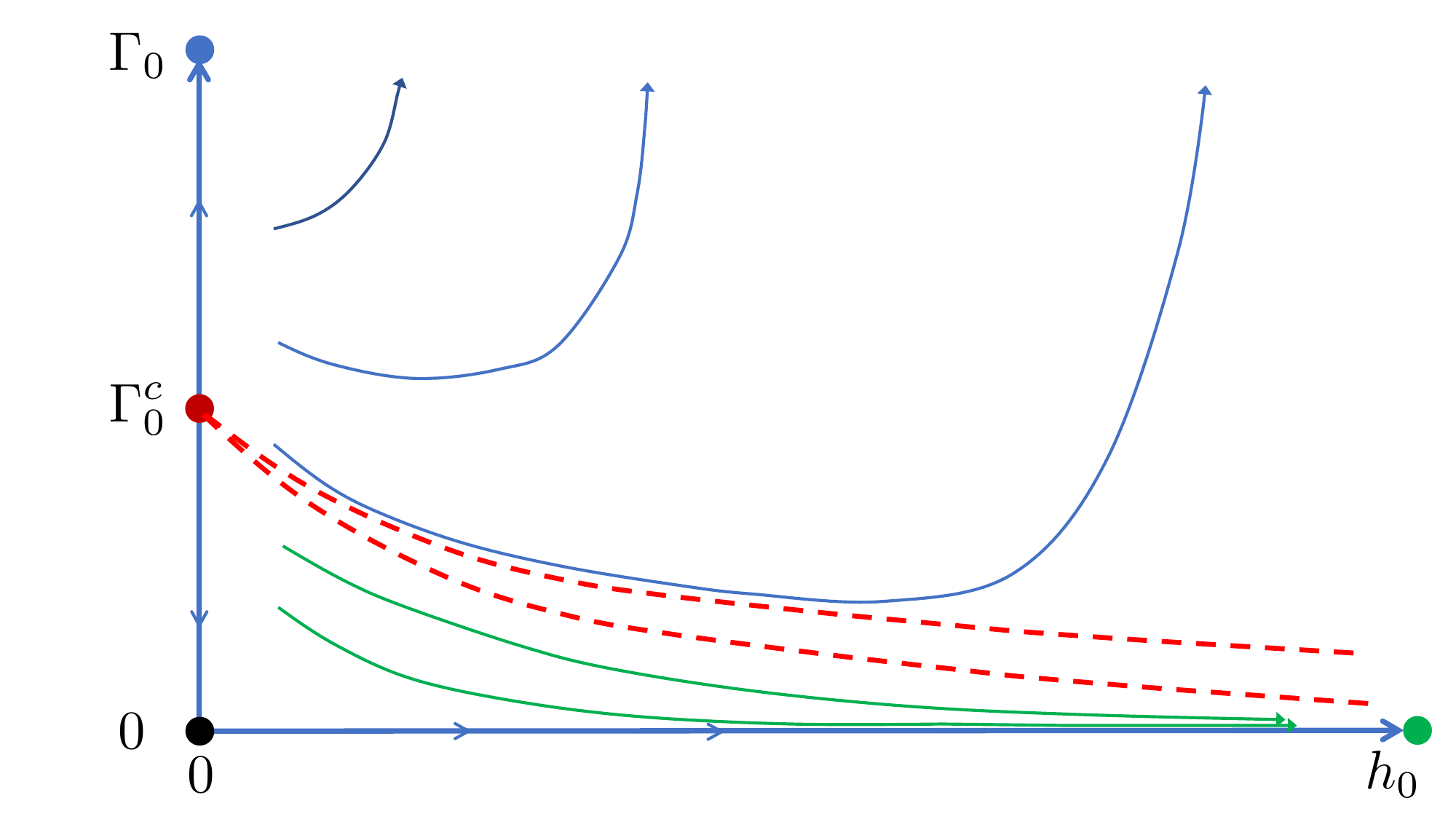}
	\vskip -0.3cm
\caption{Schematic RG phase-diagram of the model in Eq.(\ref{Hamilton}) in the thermodynamic limit using the parameters of the box-like distribution in Eq.(\ref{eq:J_distrib}). At $h_0=0$ there are two trivial fixed-points: one at $\Gamma_0=0$ and indicated by a black circle which controls he quantum ordered phase and another at $\Gamma_0 \to \infty$ and indicated by a blue circle which controls he quantum disordered phase. These are separated by a non-trivial IDFP denoted by a red circle and located at $\Gamma_0^c$. In the classical limit $\Gamma_0=0$, we have the random-field Ising model, which has a classical disordered phase for any value of $h_0>0$ and its properties are controlled by a fixed-point at  $h_0 \to \infty$ and indicated by a green circle. For general values of the parameters the system has no long-range order \textcolor{black}{and the RG trajectories are attracted by a set of disordered fixed points, in which the couplings are negligible and the disorder is either due to random quantum fluctuations (illustrated by blue trajectories), or due to classical random field effects (illustrated by green trajectories).  Between the two regimes there is a cross-over region, the borders of which are illustrated by dashed red lines and which are denoted by $\Gamma_s^+(h_0)>\Gamma_s^-(h_0)$. In the vicinity of the IDFP the cross-over region is sharp,  $\lim_{h_0 \to 0} [\Gamma_s^+(h_0)-\Gamma_s^-(h_0)]=0$ and the local separatrix defines the relevant scaling direction.}}
\label{fig_1}	
\end{figure} 
%%%%%%%%%%%%%%%%%%%%%

A short report about our preliminary investigations of this model has been published in Ref.\cite{Pet__2023}. In the present paper, we go beyond the results in Ref.\cite{Pet__2023} in several aspects. Here, we study the location of the \textcolor{black}{cross-over region}, which indicates the relevant scaling direction of the IDFP at $h_0 \to 0$ and calculate the value of the correlation-length critical exponent. In the numerical calculations, we use ten-times more samples in order to reduce the statistical error. We also study the effect of the dilution parameter, $\zeta$ on the value of the critical exponents and study the distribution of the low-energy excitations and compare it with the form of extreme-value statistics. We would like to point out that for the sake of clarity we repeat some technical aspects of the methodology that are necessary for a better understanding.

The rest of our paper is organized as follows.
In Sec.\ref{sec:SDRG} the SDRG method is introduced and its fixed-points are analysed. Numerical results for finite random longitudinal fields are presented in Sec.\ref{sec:results} and discussed in Sec.\ref{sec:disc}.

\section{SDRG treatment}
\label{sec:SDRG}

In the SDRG method\cite{IGLOI2005277,Igloi2018} we consider local parameters in the Hamiltonian in Eq.(\ref{Hamilton}). At position $i$, these are couplings, having a value $J_i$, or sites, having the characteristic parameter: 
\be
\gamma_i=\sqrt{\Gamma_i^2+h_i^2}\;.
\label{epsilon_i}
\ee 
The largest value of the corresponding gap, denoted by $\Omega$, sets the energy-scale in the problem, and this parameter is eliminated. At the same time, new terms in the Hamiltonian are generated through perturbation calculation between the remaining degrees of freedom. After successive iteration of the procedure, $\Omega$ will approach the fixed-point, with $\Omega^*=0$, where 
one makes an analysis of the distribution of the different parameters and calculates the scaling properties. For the Hamiltonian in Eq.(\ref{Hamilton}), there are two elementary decimation steps, which are illustrated in Fig.\ref{fig_2}\cite{Pet__2023}.

%\subsection{Elementary decimation steps}
%\label{sec:decimation}
%%%%%%%%%%%%%%%%%%%%%%%%% Fig 0 %%%%%%%%%%
\begin{figure}[h!]
\center
\includegraphics[width=1.\columnwidth]{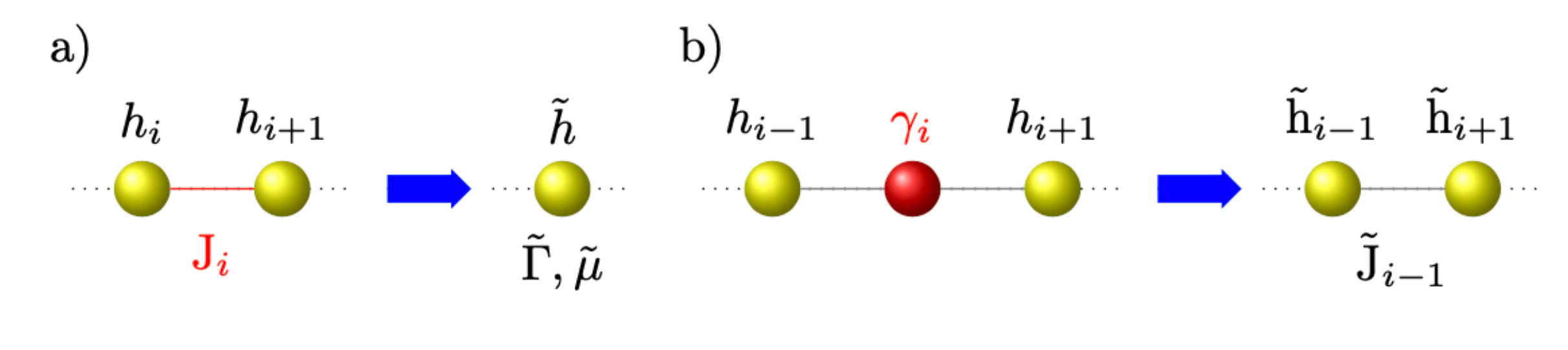}
	\vskip -0.4cm
\caption{Illustration of the SDRG decimation steps a) for strong coupling and b) for strong field decimation. The decimated parameters are denoted by red colour.}
\label{fig_2}	
\end{figure} 
%%%%%%%%%%%%%%%%%%%%%

%\subsection{Strong-coupling decimation}

If the largest local term in the Hamiltonian is a coupling, corresponding to a gap of $\Omega=2J_i$, connecting sites $i$ and $i+1$, then these two sites will be merged to a spin cluster in the presence of a (renormalized) transverse field $\tilde{\Gamma}$ and a longitudinal field $\tilde{h}$ . The magnetic moment of the cluster is then given by: $\tilde{\mu}=\mu_i+\mu_{i+1}$, with the initial magnetic moments $\mu_i=\mu_{i+1}=1$. In second-order perturbation calculation we obtain for the renormalized parameters:
\be
\tilde{\Gamma}=\frac{\Gamma_i\Gamma_{i+1}}{J_i},\quad \tilde{h}=h_i+h_{i+1}\;.
\label{tilde_Gamma_h}
\ee

%\subsection{Strong field decimation}
%\label{sec:strong_field}

If the largest local term in the Hamiltonian is related to a site $i$, and the associated parameter is $\gamma_i$, then this site will be eliminated, but the longitudinal magnetic field, $h_i$, will be transformed at the remaining neighbouring sites. The new renormalized coupling between the remaining sites $i-1$ and $i+1$ can be calculated from the energy levels with fixed spins at these sites. Denoting by $s_{i\pm1}=+$ ($-$) a $\uparrow$ ($\downarrow$) boundary state, the eigenvalue problem with different boundary conditions has the lowest energy as:
\begin{equation}
E_{s_{i-1},s_{i+1}}=-\sqrt{\Gamma_i^2+(s_{i-1}J_{i-1}+s_{i+1}J_i+h_i)^2}\;.
\end{equation}
The renormalised coupling is given by:
\begin{equation}
\tilde{J}=-(E_{\uparrow \uparrow}+E_{\downarrow \downarrow}-E_{\uparrow \downarrow}-E_{\downarrow \uparrow})/4 \approx \frac{J_{i-1}J_i}{\gamma_i}\left(\frac{ \Gamma_i}{\gamma_i}\right)^{2}\;,
\label{tilde_J}
\end{equation}
where the last relation is calculated perturbatively.

For the excess longitudinal fields we have:
\be
\Delta h_{i-1}=-(E_{\uparrow \uparrow}-E_{\downarrow \downarrow}+E_{\uparrow \downarrow}-E_{\downarrow \uparrow})/4 \approx \frac{J_{i-1}h_i}{\gamma_i}\;,
\ee
and
\be
\Delta h_{i+1}=-(E_{\uparrow \uparrow}-E_{\downarrow \downarrow}-E_{\uparrow \downarrow}+E_{\downarrow \uparrow})/4 \approx \frac{J_{i}h_i}{\gamma_i}\;,
\ee
so that 
\be
\tilde{h}_{i\pm1}={h}_{i\pm1}+\Delta {h}_{i\pm1}\;.
\ee
We note that in the absence of longitudinal fields, $h_i=0$, when $\gamma_i=\Gamma_i$, the decimation equation in Eq.(\ref{tilde_Gamma_h}) can be written in a non-perturbative way:
\begin{align}
&\tilde{\Gamma}=\left[\sqrt{J_i^2+(\Gamma_i+\Gamma_{i+1})^2}-\sqrt{J_i^2+(\Gamma_i-\Gamma_{i+1})^2}\right]/2,\nonumber \\
&\tilde{h}=0\;.
\end{align}
In the numerical calculations, we generally used the non-perturbative expressions for the renormalized parameters in order to keep the iterations more stable.

\subsection{SDRG fixed-points}
\label{fixed_points}
Here, we interpret the fixed-points already announced in the phase-diagram in Fig.\ref{fig_1}. The fixed-points at $h_0=0$ are those of the random transverse-field Ising chain, the properties of which are known through the solution of the SDRG equations\cite{PhysRevLett.69.534,PhysRevB.51.6411}.  The trivial fixed-points of the transformation are at $\Gamma_0=0$ (controlling the ordered phase) and  at $\Gamma_0 \to \infty$ (controlling the quantum disordered phase).  The non-trivial fixed-point, which governs the critical behaviour is located at $\Gamma_0=\Gamma_0^c$ and it is an IDFP. 

%%%%%%%%%%%%%%%%%%%%%%%%% Fig 1a %%%%%%%%%%
\begin{figure}[h!]
\includegraphics[width=1.\columnwidth]{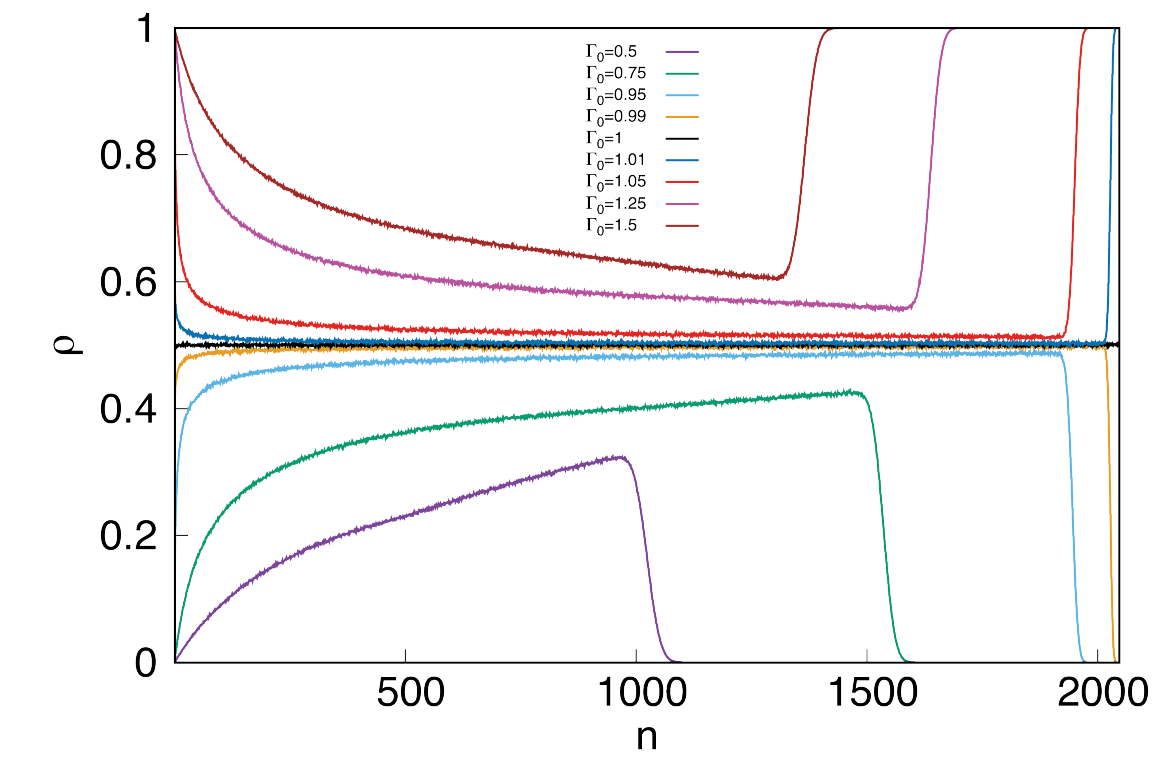}
\includegraphics[width=1.\columnwidth]{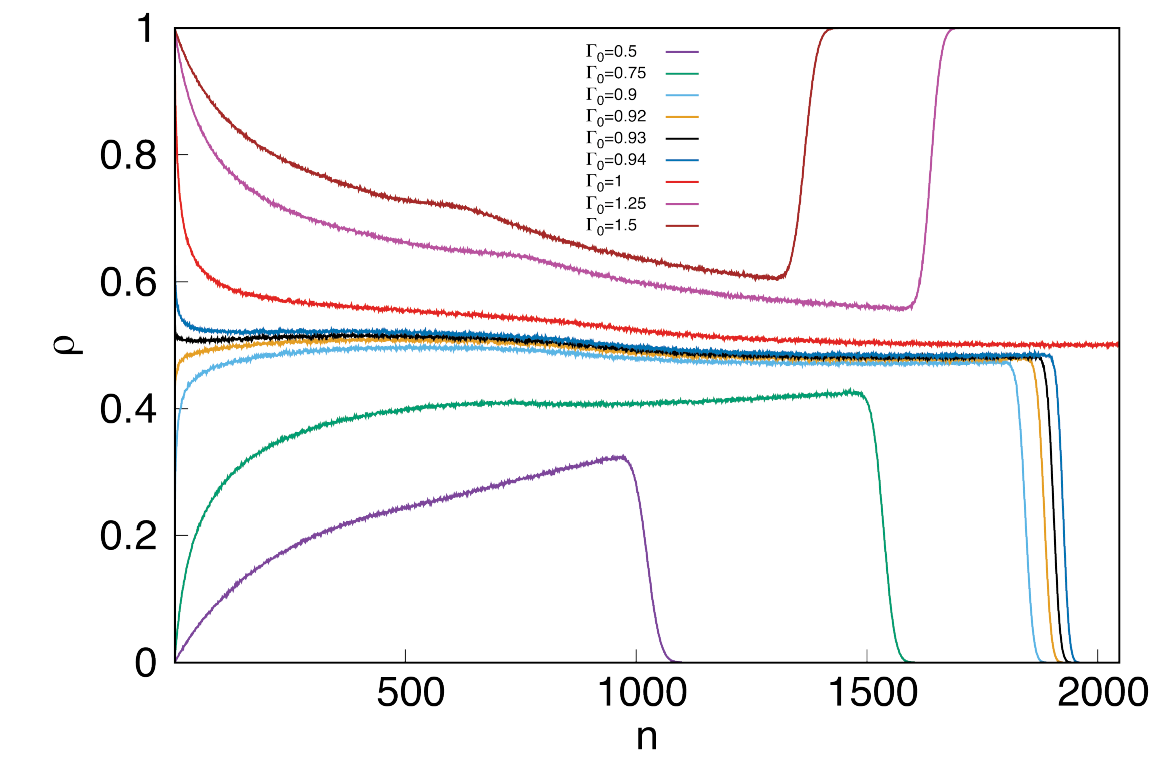}
	\vskip -0.cm
\caption{Fraction of performed site (or field) decimations during the SDRG process, \textcolor{black}{$\rho$} as a function of remaining sites, $n$, for different strengths of the transverse field, $\Gamma_0$ at $h_0=0$. The results are obtained on a chain with $L=2048$ and the average is made over $100\,000$ random samples. While the second order approximation of the decimation steps leads to an IDFP at $\Gamma_0^c=1$ (upper panel), the more detailed decimation rules followed here are asymmetric, shifting the IDFP slightly to $\Gamma_0^c\approx 0.93$ (lower panel), without affecting the universal behaviour. \textcolor{black}{The abrupt changes in the trajectories are due to the form of the disorder in Eq.(\ref{eq:J_distrib}) for large $n$ or due to the proximity to the fixed point for small $n$.}
}
\label{fig_3}	
\end{figure} 
%%%%%%%%%%%%%%%%%%%%%

The decimation process is illustrated in Fig.\ref{fig_3}, in which the fraction of site (or field) decimations, \textcolor{black}{$\rho$} are shown as a function of the number of remaining sites, $n$, for different values of the transverse field parameter, $\Gamma_0$. In the ordered phase, \textcolor{black}{for $\Gamma_0<\Gamma_0^c$,} dominantly couplings are decimated, \textcolor{black}{thus generally $\rho<0.5$}. On the contrary, in the disordered phase, \textcolor{black}{for $\Gamma_0>\Gamma_0^c$,} dominantly fields are decimated \textcolor{black}{and $\rho>0.5$}. At the critical point at $\Gamma_0=\Gamma_0^c$, the fraction of coupling- and site-decimations are the same: $\rho=0.5$. If we use the second order approximation of the decimation steps we obtain $\Gamma_0^c=1$, which follows from duality \textcolor{black}{and illustrated in the upper panel of Fig.\ref{fig_3}}. The more detailed decimation rules used here are asymmetric, shifting the IDFP to $\Gamma_0^c\approx 0.93$, \textcolor{black}{see in the lower panel of Fig.\ref{fig_3}}.

At the IDFP, the energy scale, $\epsilon$, which is the smallest gap, scales with the length $L$ as:
\be
\ln \epsilon \sim L^{\psi},\quad \psi=1/2\;.
\label{psi}
\ee

The magnetization moment, $\mu$ has a power-law $L$-dependence at the critical point:
\be
\mu \sim L^{d_f}, \quad d_f=(1+\sqrt{5})/4\;.
\label{d_f}
\ee

In the disordered phase $\delta=\Gamma_0-\Gamma_0^c>0$, the average correlations 
decay exponentially with the true correlation length:
\be
\xi \sim 1/\delta^{\nu},\quad \nu=2\;.
\label{nu}
\ee
We mention that the decay of the typical correlations involves a different exponent:
\be
\nu_{\rm typ}=1\;.
\label{nu_typ}
\ee
Close to the critical point in the disordered phase, in the so called Griffiths phase, the energy-scale goes to zero as:
\be
\epsilon \sim L^{-z}\;,
\label{L_z}
\ee
where $z$ is the dynamical exponent, which also can be calculated exactly\cite{PhysRevE.58.4238,PhysRevLett.86.1343,PhysRevB.65.064416}.

%\subsection{Classical random-field Ising chain at $\Gamma_0=0$}
Another trivial fixed-point of the SDRG transformation is located at $\Gamma_0=0$ and $h_0 \to \infty$, and controls the properties of the classical random-field Ising chain.
It is known rigorously that in the classical random-field Ising model there is no ferromagnetic order in dimensions $d \le 2$\cite{PhysRevLett.35.1399,PhysRevLett.59.1829,Binder1983}. Consequently, in our model in $d=1$ the system is classically disordered for any value of $h_0>0$. This result follows also from the SDRG equations in Sec.\ref{sec:SDRG}. Having a small random-field parameter, $h_0 \ll 1$, in the first steps of the renormalization typically couplings are decimated. After eliminating a fraction of $s$ couplings, composite spins with a typical linear size, $\ell$ and moment $\ell \sim \tilde{\mu} \sim 1/s$ are created, having typical longitudinal fields as $\tilde{h} \sim h_0/\sqrt{s}$. When $\tilde{h}$ exceeds the value of the typical couplings, which happens at  $h_0^2>s \sim 1/\ell$, typically fields are decimated, which will result in a set of separated spin clusters, since the couplings between those will be vanishing, in accordance with Eq.(\ref{tilde_J}). The correlation length in the system, $\xi(h_0)$, is related to the linear extension of the disconnected clusters:
\be
\xi(h_0) \sim \ell \sim \frac{1}{h_0^2}\;,
\ee
in agreement with exact results\cite{PhysRevB.27.4503,FIgloi_1994}. The trivial fixed-point, which describes the behaviour of the disordered classical random-field Ising model, is located at $h_0 \to \infty$ and indicated by a green circle in Fig.\ref{fig_1}.

\section{Numerical study for $h_0>0$ and $\Gamma_0>0$}
\label{sec:results}
In this section, we turn on both the random transverse and the random longitudinal fields and study the behaviour of the renormalization flow. This way, we aim to explore the \textit{terra incognita} in Fig.\ref{fig_1}. We aim also to determine the scaling properties of the non-trivial IDFP in the simultaneous presence of random couplings and random transverse and longitudinal fields.

 \subsection{Properties of the RG-flow}
\label{sec:RG_flow}

 %%%%%%%%%%%%%%%%%%%%%%%%% Fig 1a %%%%%%%%%%
%\begin{figure}[h!]
%\includegraphics[width=1\columnwidth]{Fig_casestudy05.eps}
%\includegraphics[width=1.\columnwidth]{Fig2.pdf}
%\includegraphics[width=1\columnwidth]{Fig_casestudy15.eps}
%	\vskip -0.5cm
%\caption{Renormalized value of the log-energy excitations as a function of remaining sites, $n$, for the non-perturbed model with $h_0=0$ (green symbols) and for the model with a random longitudinal field ($\ln h_0=-6$, blue symbols). The value of the log random field at a site decimation is shown by yellow symbols. The results are obtained on a chain with $L=2048$ and the log-variables are averaged over $10000$ random samples. Upper panel: starting from the ordered nonperturbed phase $\Gamma_0=0.5$; second panel: starting from the expected separation point: $\Gamma_0=\Gamma_s=0.851$; third panel: starting from the nonperturbed critical point $\Gamma_0=1.$; lower panel:  starting from the disordered nonperturbed phase $\Gamma_0=1.5$. The horizontal line at $\ln h_0=-6$ shows the parameter of the original distribution of the random longitudinal fields.}
%\label{fig_1a}	
%\end{figure} 
%%%%%%%%%%%%%%%%%%%%%

Key information about the renormalization process can be obtained from an analysis of the fraction of site (and/or bond) decimations, $\rho$ versus the number of remaining sites, $n$, which is illustrated in \textcolor{black}{the upper panel of} Fig.\ref{fig_4} for different values of $\Gamma_0$, and at a finite value of the longitudinal field, $\ln(h_0)=-6$. This is to be compared with a similar analysis performed at $h_0=0$ and presented in Fig.\ref{fig_3}. \textcolor{black}{Up to $n>n^*(\Gamma_0)$ the curves are indistinguishable, their difference, $\Delta \rho$ is shown in the lower panel of Fig.\ref{fig_4}}.  The properties of the RG-flows are different for larger values of $\Gamma_0>\Gamma_s^+$ from that obtained at relatively smaller values, $\Gamma_0<\Gamma_s^-$. In the first regime  \textcolor{black}{we have $\Delta \rho<\Delta \rho^* \sim 0.5$, for $\forall n$, so that} the RG-flow is very much similar to that in Fig.\ref{fig_3}, and the system renormalises to a quantum disordered phase. \textcolor{black}{We have generally for the deviation point in the lower panel of Fig.\ref{fig_4} $n^*(\Gamma_0)<n^*(\Gamma_s^+)$ for $\Gamma_0>\Gamma_s^+$.} If we start with $\Gamma_0<\Gamma_s^-$ the RG-flow is similar to that for $h_0=0$ only in the initial period, in which dominantly couplings are decimated. If the number of remaining sites is less than a limit, $n < n^*(\Gamma_0)$, than the renormalised longitudinal fields are dominantly decimated, and the system renormalises to a classical random-field Ising chain. \textcolor{black}{Here we have $n^*(\Gamma_0)<n^*(\Gamma_s^-)$ for $\Gamma_0<\Gamma_s^-$. The actual values of $\Gamma_s^{\pm}$ will be defined in Sec.\ref{sec:separatrix}, here we can say that $n^*(\Gamma_s^+) \approx n^*(\Gamma_s^-)$.}

%%%%%%%%%%%%%%%%%%%%%%%%% Fig 1a %%%%%%%%%%
\begin{figure}[h!]
\includegraphics[width=1.\columnwidth]{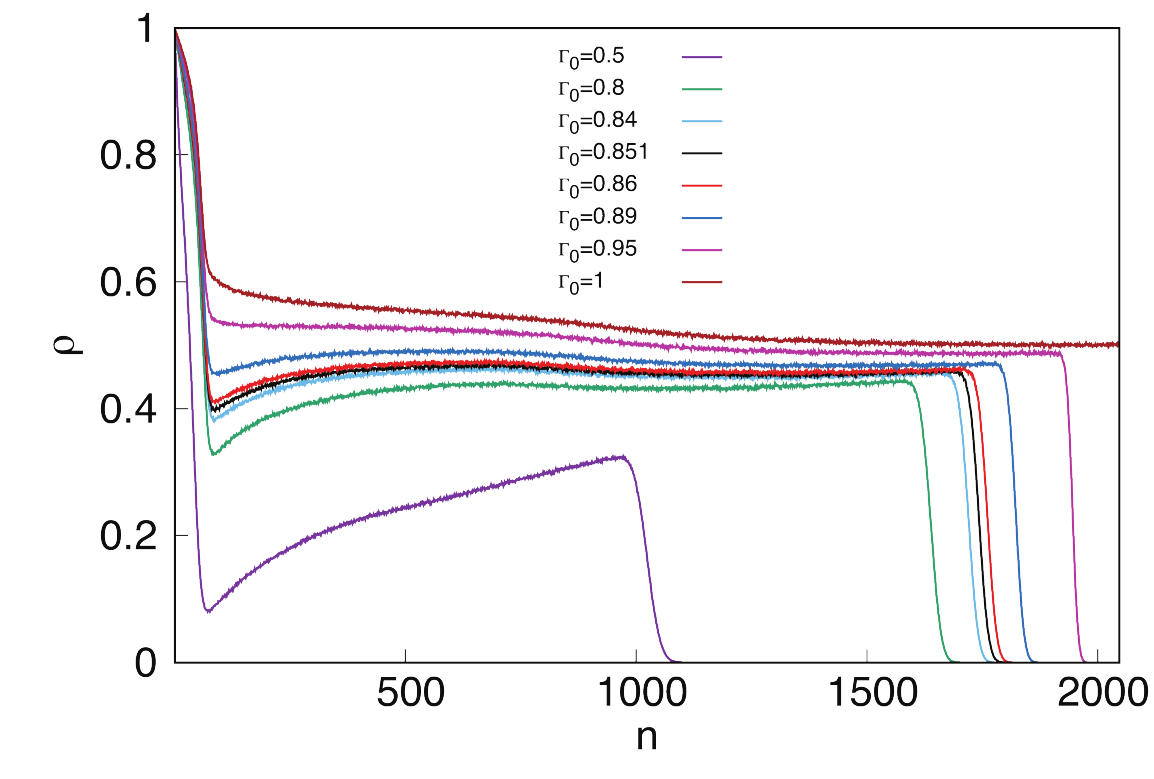}
\includegraphics[width=1.\columnwidth]{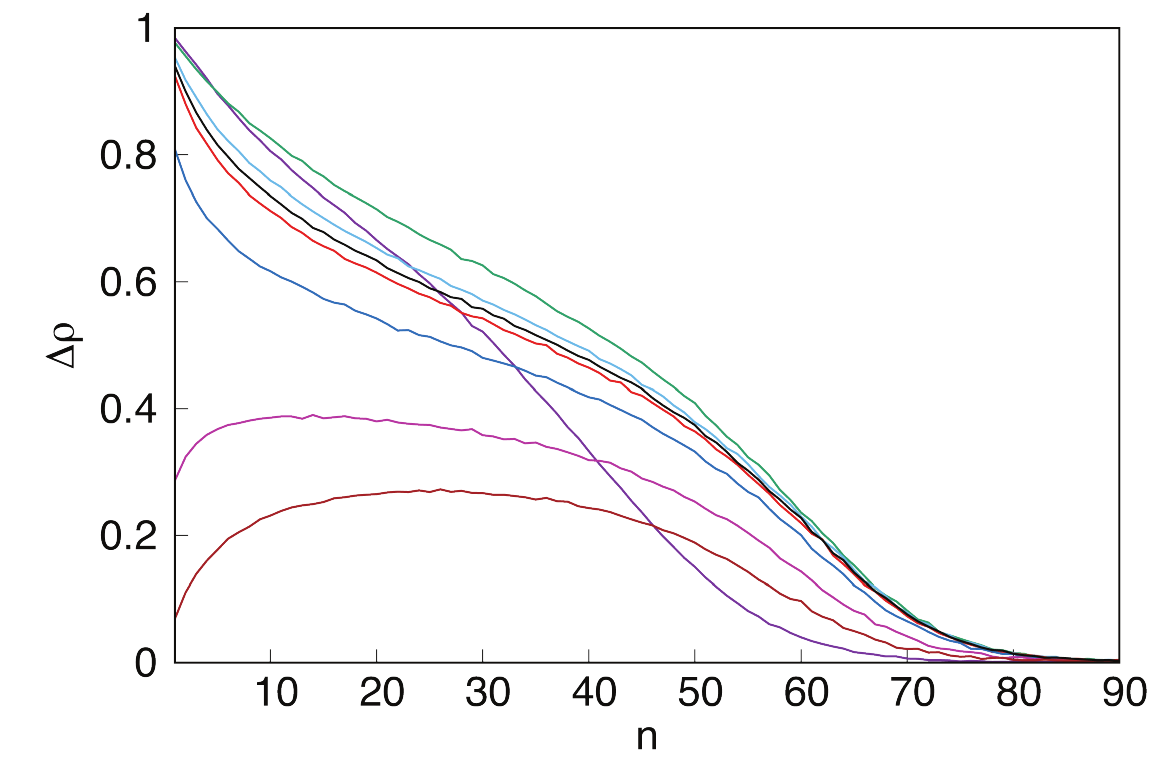}
	\vskip -0.cm
\caption{Upper panel: fraction of performed site (or field) decimations during the SDRG process as a function of remaining sites, $n$, for different strengths of the transverse field, $\Gamma_0$, at a random longitudinal field $\ln h_0=-6$. The results are obtained on a chain with $L=2048$ and the average is made over $100\,000$ random samples.
Lower panel: difference between the $\rho$ values calculated at $\ln h_0=-6$ and at $h_0=0.$}
\label{fig_4}	
\end{figure} 
%%%%%%%%%%%%%%%%%%%%%

Between the two regimes there is a cross-over region for $\Gamma_s^-<\Gamma_0<\Gamma_s^+$, the middle value of which \textcolor{black}{$\Gamma_s$ is characterised by the fact, that $n^*(\Gamma_s)$ has the maximum value, thus the random longitudinal field has the strongest effect to deviate the renormalization of the system from the original trajectory with $h_0=0$. In our case ($\ln h_0=-6$) it is close to $\Gamma_0\approx 0.851$. For a more accurate calculation see in Sec.\ref{sec:separatrix}.} In the cross-over region in the early starting period slightly dominantly couplings are decimated, which results in the increase of the longitudinal fields to such a value, that the combined fields, $\gamma$ and the couplings will renormalize in a symmetric fashion. In the concluding RG steps, the grown-up longitudinal fields will stop the further rapid decrease of the log excitation energy and the final state will be the result of all three parameters in the Hamiltonian. We identify $\Gamma_s^{\pm}$ as the position of the borders of the cross-over regions in Fig.\ref{fig_1}.

The point with $\Gamma_0=\Gamma_0^c$, which corresponds to the critical system at $h_0=0$, will be (slightly) above the cross-over region for $h_0>0$ and in the early starting period couplings and fields are decimated in a symmetric way, but as the longitudinal fields increase the combined fields, $\gamma$ in Eq.(\ref{epsilon_i}) will be dominant over the couplings and the system will renormalize to a quantum disordered state.

%%%%%%%%%%%%%%%%%%%%%%%%% Fig 1a %%%%%%%%%%
\begin{figure}[h!]
\vskip 0.cm
\includegraphics[width=1.\columnwidth]{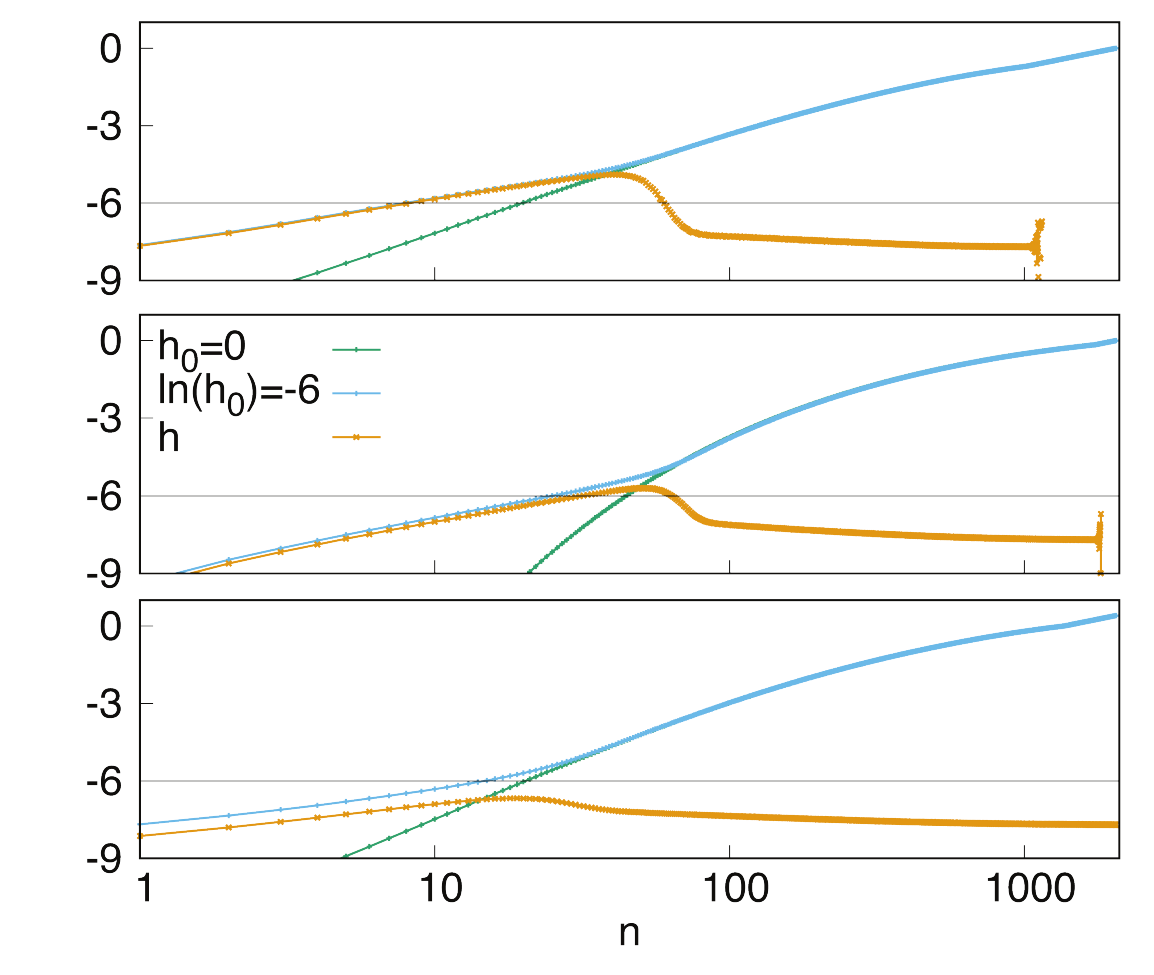}
	\vskip -0.4cm
\caption{\textcolor{black}{Renormalized value of the log-energy excitations as a function of remaining sites, $n$, for the non-perturbed model with $h_0=0$ (green symbols) and for the model with a random longitudinal field ($\ln h_0=-6$, blue symbols). The value of the log longitudinal random field absolute values at a site decimation is shown by orange symbols. The results are obtained on a chain with $L=2048$ and the log-variables are averaged over $100\,000$ random samples. Upper panel: starting from the ordered unperturbed phase $\Gamma_0=0.5$; middle panel: starting from the middle of the perturbed cross-over region $\Gamma_0=0.851$; lower panel:  starting from the disordered unperturbed phase $\Gamma_0=1.5$. The horizontal line at $\ln h_0=-6$ shows the parameter of the original distribution of the random longitudinal fields.}
}
\label{fig_4a}	
\end{figure} 
%%%%%%%%%%%%%%%%%%%%%

\textcolor{black}{In Fig.\ref{fig_4a} we compare the scaling behaviour of the log-energy excitations as a function of the remaining sites, $n$, of the non-perturbed model with $h_0=0$ and the model with random longitudinal fields ($\ln h_0=-6$). Here, we show three different points: $\Gamma_0=0.5$ - starting from the ordered unperturbed phase; $\Gamma_0=0.851$ - the middle of the perturbed cross-over region; $\Gamma_0=1.5$ - unperturbed disordered phase. In the figure, we also present the renormalized value of the random longitudinal fields. In the starting period, when the random longitudinal fields are negligible, the two models renormalize in the same fashion, which will be changed, when the size of the random longitudinal fields will approach the value of the excitation energy. For $\Gamma_0=0.5<\Gamma_s^-$ in the starting period dominantly couplings are decimated and the renormalized transverse fields become negligible, thus the quantum fluctuations are eliminated and the system behaves as a classical one. In the concluding renormalization steps, the random longitudinal fields are dominant, and the properties of the system are controlled by the classical random-field Ising chain. On the contrary, for $\Gamma_0=1.5>\Gamma_s^+$ in the starting renormalization steps, dominantly transverse fields are decimated, the strength of the couplings is strongly reduced and at the same time the renormalized longitudinal fields will be negligible compared to the random transverse fields. In the concluding renormalization steps, the quantum fluctuations due to the random transverse fields are dominant and the properties of the system are controlled by a disordered quantum phase. At the cross-over region $\Gamma_s^-<\Gamma_0<\Gamma_s^+$, the longitudinal and transverse fields play a similar role and the excitation energy is the smallest at this point. This observation will be used to identify the value of $\Gamma_s$ in the section \ref{sec:separatrix}.}

We can thus conclude that for general values of the parameters, the system has two disordered regions, which are separated by a cross-over region, the borders of which are indicated by dashed red lines in Fig.\ref{fig_1}. This starts from the IDFP and bends downwards, due to the fact that the gap increases with increasing $h$, see in Eq.(\ref{epsilon_i}). Below the cross-over region, the RG-flows are attracted by fixed-points dominated by classical random-field Ising model effects, while above the cross-over the RG-flows scale towards a quantum disordered phase. The starting part of the cross-over at the IDFP is sharp and defines the relevant scaling direction. In the coming section, we will define a systematic method to estimate the position of the cross-over region.

%At a fixed value of the longitudinal field, $\ln h_0=-6$ we have measured the excitation energy for different values of $\Gamma_0$ and the minimal value is observed at the separatrix, see in Fig.\ref{fig_5}. 

%%%%%%%%%%%%%%%%%%%%%%%%% Fig 1a %%%%%%%%%%
%\begin{figure}[h!]
%\includegraphics[width=1. \columnwidth]{Fig_gamma.pdf}
%	\vskip -0.3cm
%\caption{The log energy gap as a function of $\Gamma_0$ at a fixed value of $\ln h_0=-6$ for finite systems of length $L=512$ and $1024$. The minimum is present at the separatrix, close to $\Gamma_0=1$.}
%\label{fig_1c}	
%\end{figure} 
%%%%%%%%%%%%%%%%%%%%%

\subsection{Estimates for the position of the cross-over region}
\label{sec:separatrix}

%%%%%%%%%%%%%%%%%%%%%%%%% Fig 1a %%%%%%%%%%
\begin{figure}[h!]
\vskip 0.cm
\includegraphics[width=1.\columnwidth]{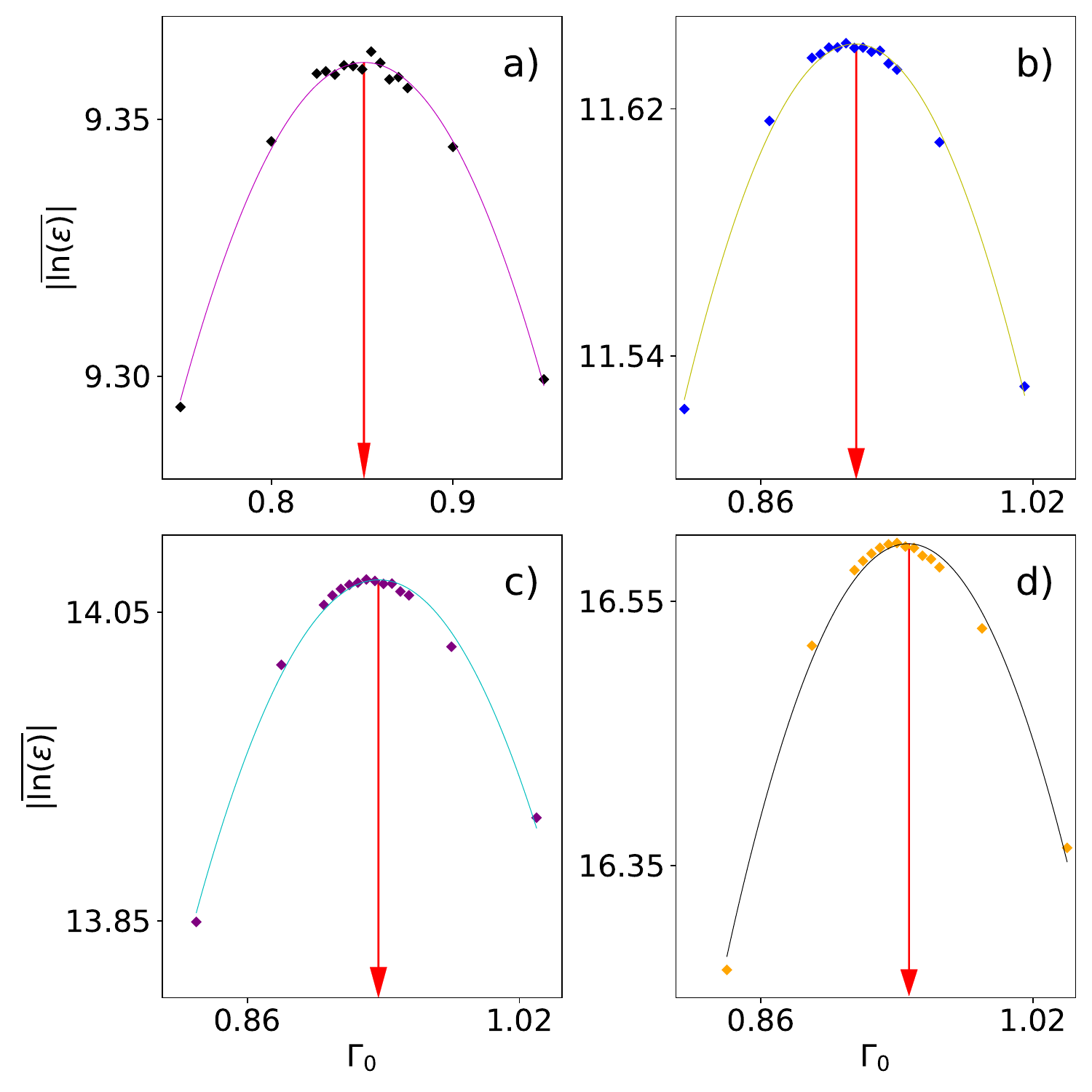}
	\vskip -0.4cm
\caption{The average of the absolute value of the log-excitation energy, $\overline{|\ln \epsilon|}$ vs.~the strength of the transverse field distribution, $\Gamma_0$, for different values of the longitudinal field:
$\ln h_0=-6$ (top left panel), $\ln h_0=-9$ (top right panel),  $\ln h_0=-12$ (bottom left panel) and $\ln h_0=-15$ (bottom right panel) at a chain length $L=2048$. The fitted parabolas are also shown. The curves exhibit a maximum at $\Gamma_s(h_0)$, which is indicated by an arrow and given by $0.851,~0.916,~0.937$ and $0.947$, for the panels in the previous order.
}
\label{fig_5}	
\end{figure} 
%%%%%%%%%%%%%%%%%%%%%

 %Studying the RG-flow at different parameters we noticed that its behaviour is qualitatively different, if the starting point is close to the ordered non-perturbed (i.e. with $h_0=0$) phase or it is close to the disordered non-perturbed phase. Between the two regimes we define an intermediate point with $\Gamma_0=\Gamma_s$, which is called the separation point. Generally $\Gamma_s$ depends on $h_0$ and as $h_0 \to 0$ then $\Gamma_s$ is expected to approach the critical value $\Gamma_0=1$ from below in the thermodynamic limit. 
  %%%%%%%%%%%%%%%%%%%%%%%%% Fig 1a %%%%%%%%%%
\begin{figure}[h!]
\includegraphics[width=1.\columnwidth]{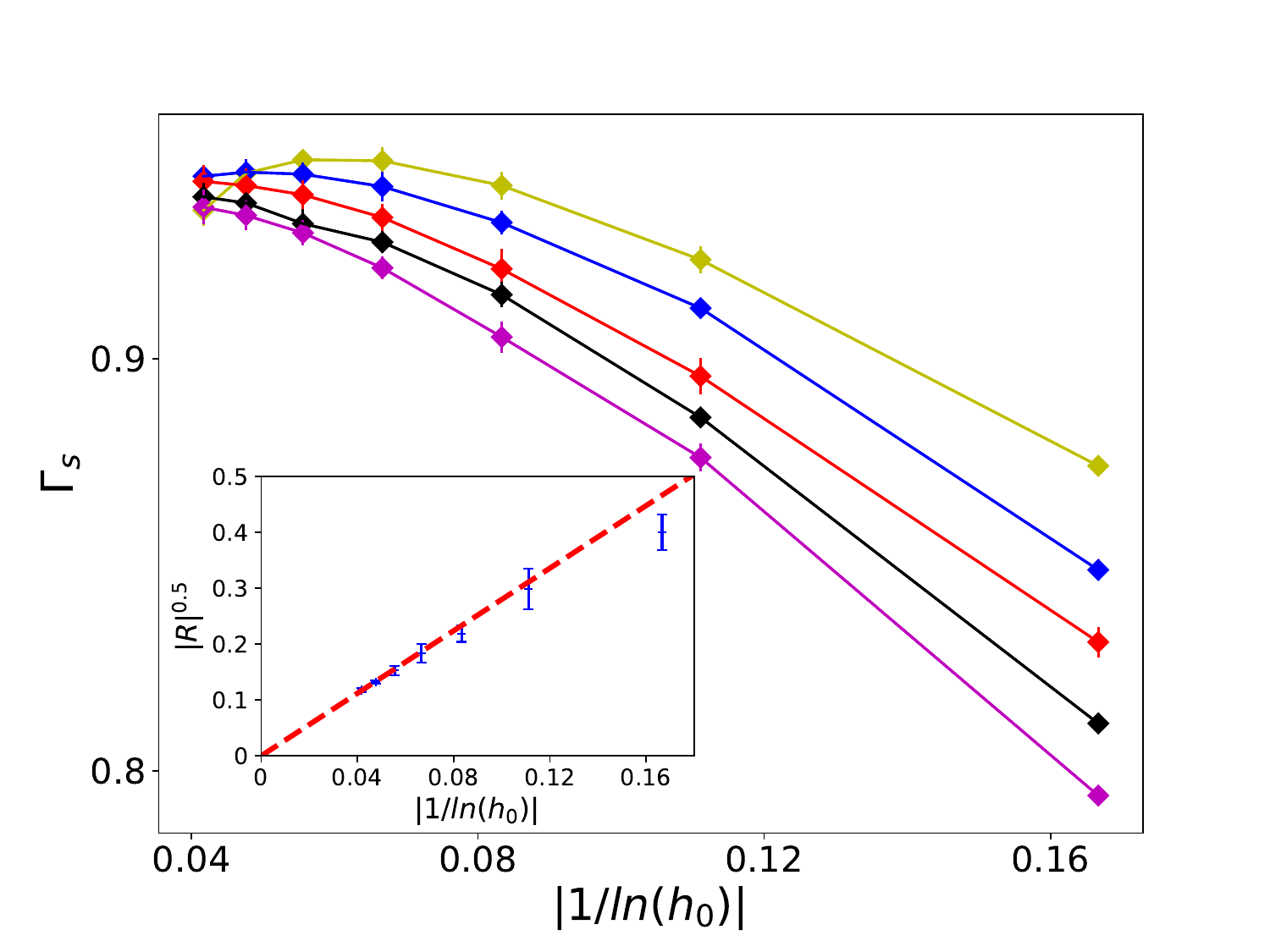}
	\vskip -0.4cm
\caption{The middle of the cross-over region $\Gamma_s(h_0)$, defined through the position of the maximum of the curves in Fig.\ref{fig_5}. plotted as a function of $1/|\ln h_0|$ for different values of the length of the chain: $L=1024,~2048,~4096,~8192$ and $16384$, from top to bottom. \textcolor{black}{In the inset, the square root of the $R$ width of the maximum of the curves in Fig.\ref{fig_5} is plotted as a function of $1/|\ln h_0|$. $R$ is measured as the radius of curvature around the maximum.}
}
\label{fig_6}	
\end{figure} 
%%%%%%%%%%%%%%%%%%%%%

%The line of separation points is illustrated by the dashed red line in the schematic RG phase diagram in Fig.\ref{fig_1b}. 
As we explained in Sec.\ref{sec:RG_flow} in the points of the cross-over region the RG transformation in the last steps contains symmetrically decimated couplings and $\gamma$ fields. Here, we rely on this property 
to define an estimate for the position of the cross-over region. 
According to the RG-rules in Sec.\ref{sec:SDRG}, the decimation steps are connected to the value of the excitation energy, therefore we
 study the $\Gamma_0$ dependence of the excitation energy, $\epsilon$, at a fixed value of the longitudinal field, $h_0$. We noticed, that $\epsilon$, which is defined as the energy-scale in the last renormalization step, has a minimum value and alternatively $\overline{|\ln \epsilon|}$  has a maximum at the same value of the parameter of the transverse field: $\Gamma_0=\Gamma_s$. This is illustrated in Fig.\ref{fig_5}, in which we plot $\overline{|\ln \epsilon|}$ as a function of $\Gamma_0$ for different values of $h_0$. It is shown that the position of the maximum value, i.e. $\Gamma_s$, depends on $h_0$, and $\Gamma_s(h_0)$ shifts towards $\Gamma_0^c$ for smaller values of $h_0$. We argue, that $\Gamma_s(h_0)$ can be considered as an estimate for the position of the middle point of the cross-over region. Indeed, at this minimum point a cross-over between two regimes takes place: for $\Gamma_0>\Gamma_s$ dominantly field-decimation takes place, whereas for $\Gamma_0<\Gamma_s$ dominantly couplings are decimated. At $\Gamma_0=\Gamma_s$ the two processes are executed symmetrically. The position of $\Gamma_s$ obtained from the analysis in Fig.\ref{fig_5} for $\ln(h_0)=-6$ is in accordance with the analysis of the decimation process shown in Fig.\ref{fig_4}. Performing the analysis shown in Fig.\ref{fig_5} for several values of $h_0$ we have obtained a set of values $\Gamma_s(h_0)$ for a given length of the chain, $L$. \textcolor{black}{We have measured the width of the maximum of the curves as the radius of curvature, $R$, so that we can define the borders of the cross-over regions as $\Gamma_s^{\pm}=\Gamma_s \pm R/2$. The measured values of $R$ averaged over the largest sizes, $L\geq2048$ is plotted in the inset of Fig.\ref{fig_6}. We noticed that the relation $\Delta \Gamma_0 \equiv R \propto 1/|\ln h_0|^2$ is valid within the statistical error for large values of $L$, thus the cross-over is indeed sharp in the limit of a small longitudinal field.} 

 Repeating the calculations for different lengths we have obtained a set of curves, shown in Fig.\ref{fig_6} as a function of $1/|\ln(h_0)|$. It is shown that for not too small values of $h_0$ the curves for a given length monotonously decrease with increasing $h_0$. A rough extrapolation 
  of this part of the curves to $h_0=0$ would result in a value, which is close to the IDFP: $\Gamma_0=\Gamma_0^c$. 
  For smaller values of $h_0$, however, the points of the curves start to bend down, which we attribute to finite-size effects, especially visible for $L=1024$. 
  This cross-over point is close to the limiting point, $\tilde{h}_0(L)$, which is identified in Sec.\ref{sec:case1}. The part of the curves not affected by finite-size effects shows a monotonically decreasing trend with increasing values of $L$. We use this part of the curves to define the relevant scaling direction, which is $L$-dependent.

\subsection{Scaling behaviour in the vicinity of the IDFP}
\label{sec:Num_results}

In this section, we study numerically the properties of the system in the vicinity of the IDFP, considering two different trajectories starting from the IDFP, considering $10^6$ random samples. 

i) In the first case, we fix the value of $\Gamma_0^c=0.93$ and consider a set of points with $h_0>0$. In this case, the 
$\Gamma_0$
coordinates of the starting point of the renormalization do not depend on the length of the chain. 
We note that preliminary results of this type of analysis has been announced in Ref.\cite{Pet__2023}. %Here, we have increased the number of random samples from $10^5$ to $10^6$.

ii) In the second case, we follow the position of the special points, $\Gamma_s$, as they are determined in Sec.\ref{sec:separatrix}. In this case, at a fixed value of $h_0$, the starting point of the renormalization is (weakly) size dependent. This size-dependence could result in differences in the critical exponents, if these are calculated through finite-size scaling.

In the numerical analysis, we used finite periodic chains with lengths $L=2^n$, $n=7,8\dots,14$ and monitored the behaviour of the system at small values of $h_0$. %At each point $10^6$ random samples are considered. 
Our numerical algorithm works in linear time as a function of $L$ with some logarithmic correction. 
At each decimation step, the local term corresponding to the maximal gap is considered, selected via using a binary heap data structure.
We have measured the average value of the log-gap, $\overline{\ln \epsilon}$, where $\epsilon$ is given by the last decimated site value: $\epsilon=\sqrt{\tilde{\Gamma}^2+\tilde{h}^2}$ and the average value of the magnetization moment, $\overline{\mu}$.
%todo: make sure the use of gamma and epsilon is consistent in the text
%These large scale calculations also convinced us, that the random longitudinal fields for $h_0>0$ represent a relevant perturbation, for large enough chains the system renormalizes to the disordered phase. 

%%%%%%%%%%%%%%%%%%%%%%%%% Fig 1 %%%%%%%%%%
\begin{figure}[h!]
\includegraphics[width=1\columnwidth]{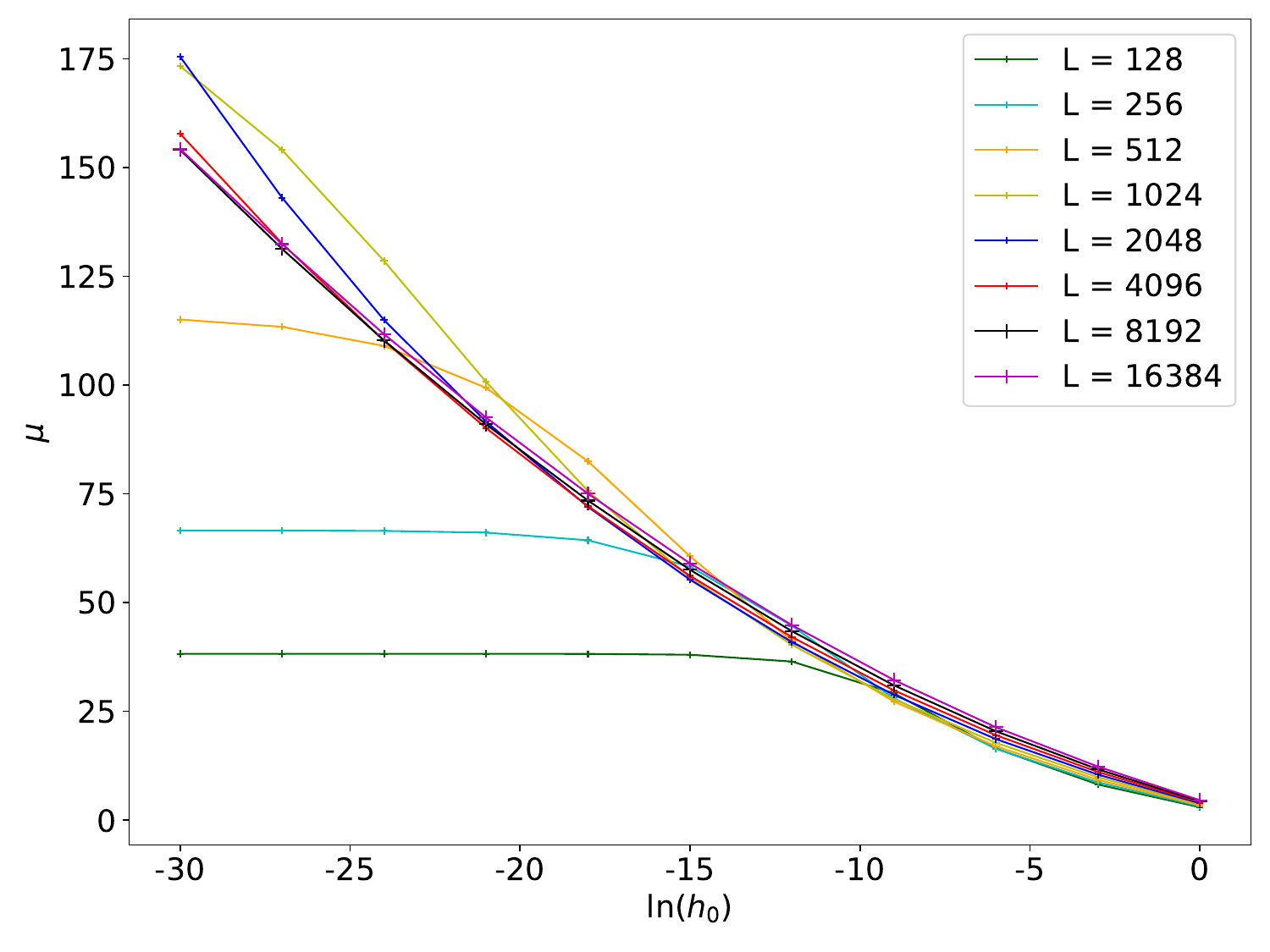}
\includegraphics[width=1\columnwidth]{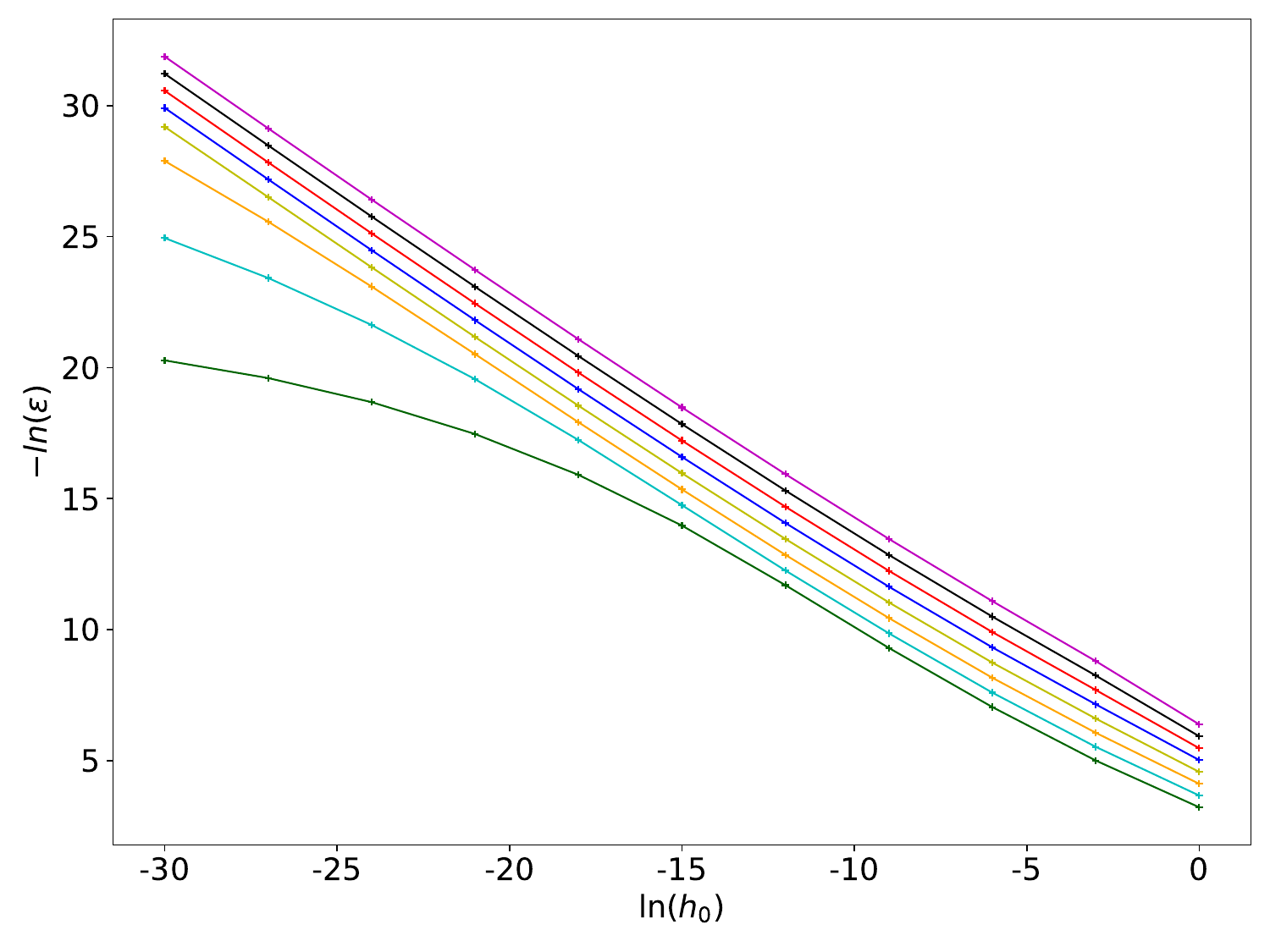}
	\vskip -0.4cm
\caption{Average magnetization moment (upper panel) and average log-gap (lower panel) as a function of $\ln h_0$ for different lengths of the chain, calculated at $\Gamma_0^c=0.93$.}
\label{fig_7}	
\end{figure} 
%%%%%%%%%%%%%%%%%%%%%

\subsubsection{Analysis along the line $\Gamma_0^c=0.93$}
\label{sec:case1}

Here, we considered a set of points with $-\ln h_0=0,3,6,\dots,30$ and the obtained results are presented in Fig.\ref{fig_7}. As shown in this figure, at a finite length, $L$, there is a cross-over behaviour if the longitudinal field is around $h_0 \approx \tilde{h}_0(L)$. For $h_0 < \tilde{h}_0(L)$ the influence of the original fixed-point at $h_0=0$ becomes dominant, so that the true asymptotic behaviour is seen only for $h_0 > \tilde{h}_0(L)$. Estimates for $\tilde{h}_0(L)$ can be obtained from the position of the inflection points in Fig.\ref{fig_7}. Equivalently, for a fixed value of  $h_0$, the length of the chain should be sufficiently large, $L  > \tilde{L}(h_0)$, in order to see the asymptotic behaviour. Deep in the asymptotic regime, the average quantities are approximately linear with $\ln h_0$ and we have the relations:
\begin{align}
\overline{\mu}_L(h_0^{(2)})-\overline{\mu}_L(h_0^{(1)}) &\approx -\kappa \ln(h_0^{(2)}/h_0^{(1)})\nonumber \\
\overline{\ln \epsilon}_L(h_0^{(2)})-\overline{\ln \epsilon}_L(h_0^{(1)}) &\approx \alpha \ln(h_0^{(2)}/h_0^{(1)})\;.
\label{fix_L}
\end{align}
 In Fig.\ref{fig_8} we present estimates for the prefactors, $-\kappa$ and $\alpha$.

%%%%%%%%%%%%%%%%%%%%%%%%% Fig 5 %%%%%%%%%%
\begin{figure}[t!]
\includegraphics[width=1\columnwidth]{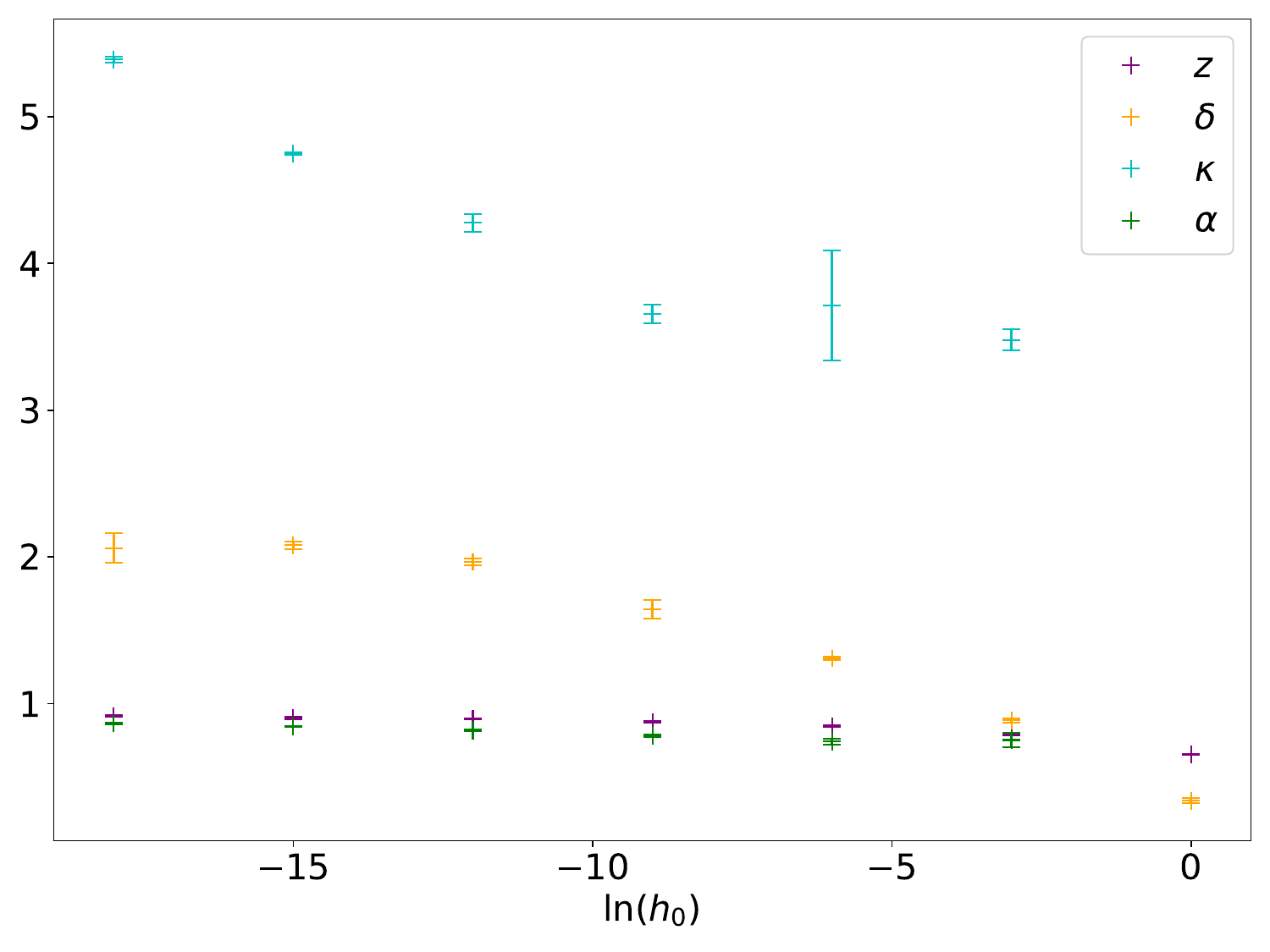}
\includegraphics[width=1\columnwidth]{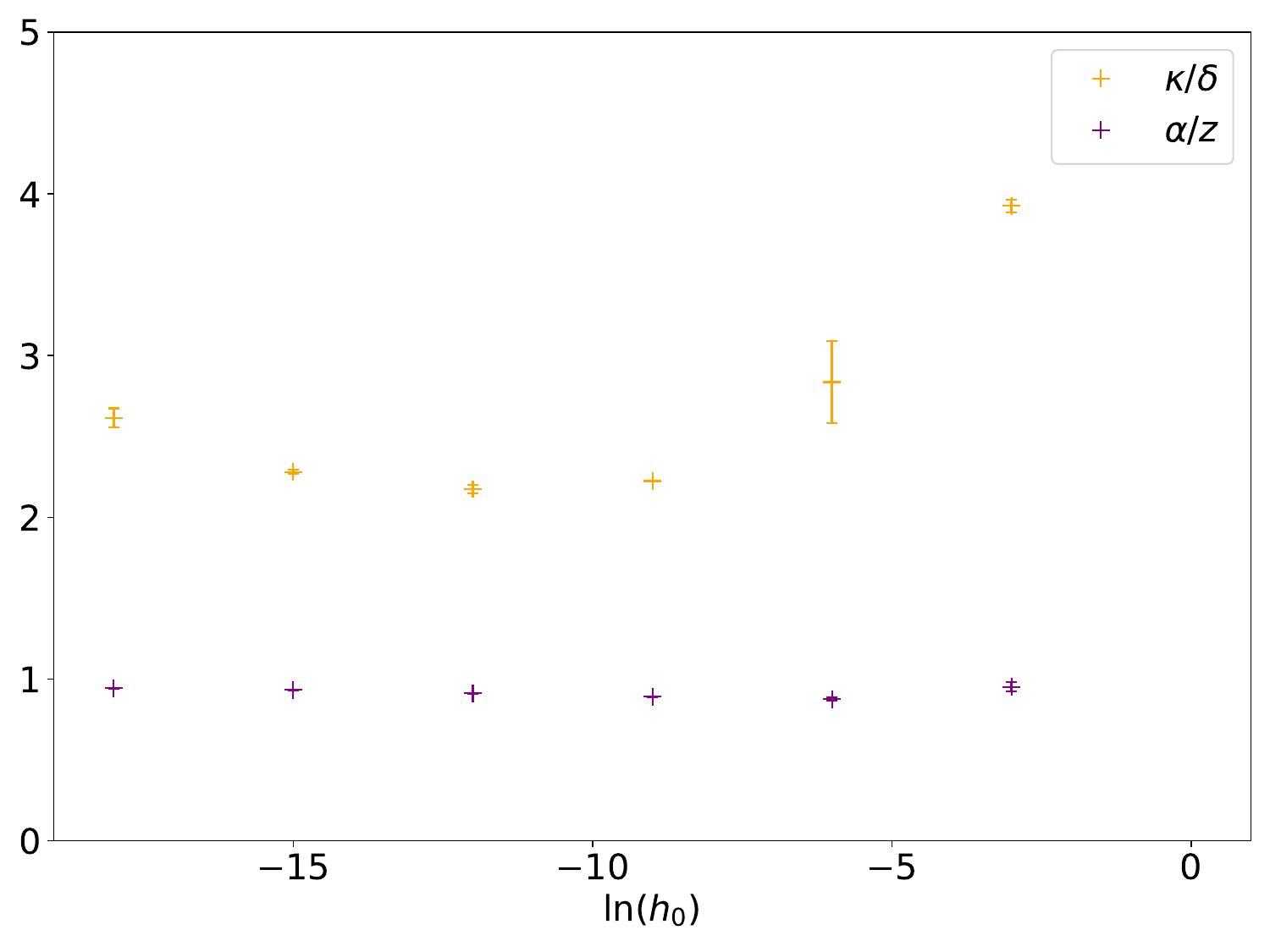}
	\vskip -0.4cm
\caption{Exponents at $\Gamma_0=0.93$. Upper panel: Estimated values of the local exponents in Eqs.(\ref{fix_L}) and (\ref{fix_h_0}) for different parameters of the distribution of the random longitudinal field, $h_0$. Lower panel: Ratio of the local exponents $\kappa/\delta$ and $\alpha/z$ for different values of $\ln h_0$. }
\label{fig_8}	
\end{figure} 
%%%%%%%%%%%%%%%%%%%%%

The curves at a fixed value of $h_0$ and for different values of $L$ are shifted in the asymptotic region. This behaviour can be summarized by the relations:
\begin{align}
\overline{\mu}_{L_2}(h_0)-\overline{\mu}_{L_1}(h_0) &\approx \delta \ln (L_2/L_1)\nonumber \\
\overline{\ln \epsilon}_{L_2}(h_0) -\overline{\ln \epsilon}_{L_1}(h_0)&\approx - z \ln (L_2/L_1)\;.
\label{fix_h_0}
\end{align}
\textcolor{black}{We mention that the second equation in Eq.(\ref{fix_h_0}) is in agreement with Eq.(\ref{L_z}), while the other relations in Eqs.(\ref{fix_L}) and (\ref{fix_h_0}) are observed only numerically.}
The estimated prefactors $\delta$ and $z$ are shown in Fig.\ref{fig_8}. \textcolor{black}{We stress that the prefactors can be used to define effective local exponents, which will approach their true values for $h_0 \to 0$, i.e. at the infinite disorder fixed point.}\textcolor{black}{We observe in Fig.\ref{fig_8}, that the parameters $\alpha$ and $z$ have only a weak $\ln h_0$ dependence, while $\kappa$ and $\delta$ show monotonous increase with increasing value of $-\ln h_0$. This latter behaviour seems to approach the scaling form at $h_0=0$, see in Eq.(\ref{d_f}).}
%%%%%%%%%%%%%%%%%%%%%%%%% Fig 2 %%%%%%%%%%
%\begin{figure}[t!]
%\includegraphics[width=1\columnwidth]{Fig_-6.pdf}
%	\vskip 0cm
%\caption{Size dependence of the average magnetization moment and the average log-gap at a fixed value of the longitudinal field $\ln h_0=-6$. The straight lines have the slopes $\delta=1.2$ and $z=0.817$, see in Eq.(\ref{fix_h_0}).}
%\label{fig_2}	
%\end{figure} 
%%%%%%%%%%%%%%%%%%%%%

Using Eqs.(\ref{fix_L}) and (\ref{fix_h_0}) we can express the difference between the magnetic moments:
\be
\overline{\mu}_{L_{\mu}}(h_0) - \overline{\mu}_{L_1}(h_0^{(1)}) \approx  -\kappa \ln(h_0/h_0^{(1)}) + \delta \ln (L_{\mu}/L_1)\;,
\label{mu_rel}
\ee
and similarly for the difference between the average log-gaps:
\be
\overline{\ln \epsilon}_{L_{\epsilon}}(h_0) - \overline{\ln \epsilon}_{L_1}(h_0^{(1)}) \approx  -\kappa \ln(h_0/h_0^{(1)}) + \delta \ln (L_{\epsilon}/L_1)\;,
\label{epsilon_rel}
\ee
\textcolor{black}{The latter two equations are in accordance with the statement that $\kappa \ln(h_0)$ and $\delta \ln (L)$ have the same dimensionality.}
If the average magnetic moments in Eq.(\ref{mu_rel}) are the same, then there is a relation between the length associated to magnetic moments, $L_{\mu}$ and the distance from the fixed-point, $h_0$ as:
\be
L_{\mu} \sim h_0^{-\nu_{\mu}},\quad \nu_{\mu}=\kappa/\delta\;,
\label{nu_mu}
\ee
provided $L_1$ and $h_0^{(1)}$ are some fixed reference values. Similar analysis of the expression for the average log-gap in Eq.(\ref{epsilon_rel}) leads to the relation:
\be
L_{\epsilon} \sim h_0^{-\nu_{\epsilon}},\quad \nu_{\epsilon}=\alpha/z\;,
\label{nu_epsilon}
\ee
where $L_{\epsilon}$ is the length associated to the energy gap.

Estimates for the correlation length exponents $\nu_{\mu}=\kappa/\delta$ and $\nu_{\epsilon}=\alpha/z$ are shown in the lower panel of Fig.\ref{fig_8}. For small values of $h_0$, the estimates for $\nu_{\epsilon}$ are stable and within the error of the approximation, these are in agreement with the value $\nu_{\epsilon} \approx 1$. On the contrary, the results for $\nu_{\mu}$ contain large errors and the estimates are larger than $\nu_{\epsilon}$, for small $h_0$ being about $\nu_{\mu}/\nu_{\epsilon} \approx 2$. However, the presence of two different length scales is unusual and could be a consequence of the specific choice of the trajectory form.
%todo: "unusual"? Do we mean "different from the traditional scaling expectations"?

\subsubsection{Analysis at the relevant scaling direction}
\label{sec:case2}

Performing the RG transformation at the special points, i.e.~starting at $h_0$ and $\Gamma_0=\Gamma_s(L,h_0)$ for a chain of length $L$, the calculated average magnetizations and the average log-gaps are presented in Fig.\ref{fig_9}, which are to be compared with the results of the previous analysis in Fig.\ref{fig_7}. In the present case, the analysis is restricted to long chains, $L \ge 1024$, 
%the same length restriction also applied fro gama=1
and for limited values of $h_0$, with $-\ln h_0=6,9,\dots,24$, which approximately satisfy the relation $h_0>\tilde{h}_0(L)$, where $\tilde{h}_0(L)$ is the limiting point defined in the beginning of Sec.\ref{sec:case1}. In this range of the parameters, one expects to obtain a special point, which has only weak finite-size corrections. This assumption is indeed fulfilled for the average log-gaps for the whole range of the $h_0$ parameter. On the contrary, for the average  magnetization moment, the curves with lengths $L=1024$ and $2048$ start to deviate from the expected asymptotic behavior for small values of $h_0<\tilde{h}_0(L)$. Therefore, to perform an analysis of the magnetization data, we restrict ourselves to the three longest chains.

%%%%%%%%%%%%%%%%%%%%%%%%% Fig 1 %%%%%%%%%%
\begin{figure}[h!]
\includegraphics[width=1\columnwidth]{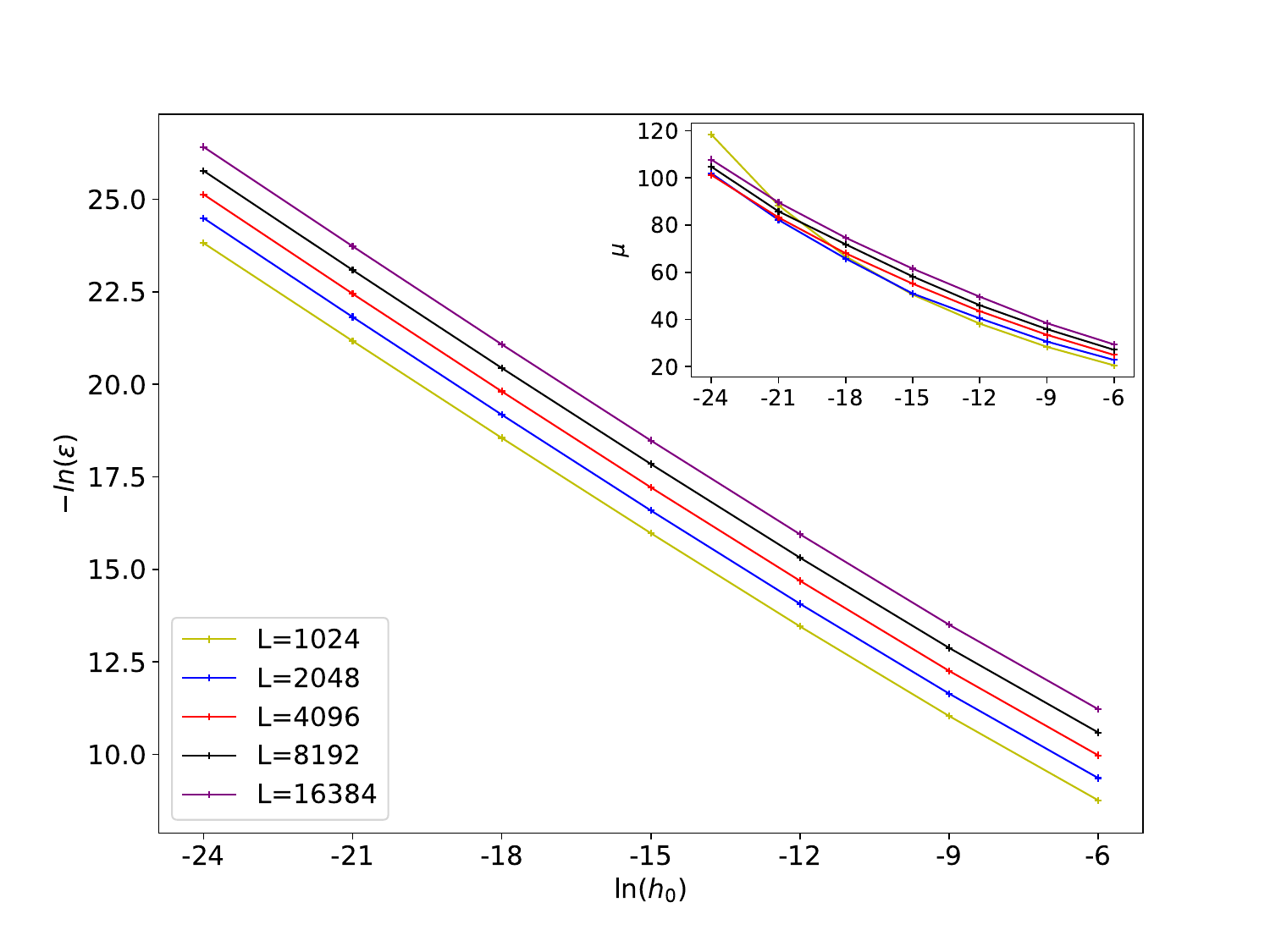}
	\vskip -0.5cm
\caption{Average log-gap (main panel) and average magnetization moment (inset) as a function of $\ln h_0$ for different lengths of the chain calculated at the separation points, which defines the relevant scaling direction.}
\label{fig_9}	
\end{figure} 
%%%%%%%%%%%%%%%%%%%%%

Analysing the data for the average log-gaps, the behaviour looks very similar to that in Sec.\ref{sec:case1}. This is also reflected in the values of the estimated exponents, $z$ and $\alpha$, which are presented in Fig.\ref{fig_10}. This observation is due to the fact that the gaps are not sensitive to small variation of the starting position of the renormalization transformation. Consequently, the correlation length exponent associated with the log-gaps in Eq.(\ref{nu_epsilon}) is given by $\nu_{\epsilon} \approx 1$. On the contrary, the data for the average magnetization moments appears to be more sensitive to the variation of the starting position. The special points are characterised by a position, $\Gamma_s(h_0,L)$, which are smaller than the value at the IDFP, $\Gamma_c=1$, resulting in a larger magnetization moment at the RG transformation. Also $\Gamma_s(h_0,L)$ have a decreasing tendency for increasing values of $L$, which is the reason of the larger values of the $\delta$ exponents, compared to those in Sec.\ref{sec:case1}. Interestingly, the curves of the average magnetization moment in Fig.\ref{fig_9} bend upwards for decreasing values of $h_0$, which will result in a set of $\kappa$ exponents, which also increase for decreasing values of $h_0$, but the ratio: $\nu_{\mu}=\kappa/\delta$ is approximately constant and can be well approximated as $\nu_{\mu} \approx 1$. We can  thus conclude that along the relevant scaling direction the correlation-length critical exponents are comparable: $\nu_{\mu} \approx \nu_{\epsilon} \equiv \nu_h$, having the value:
\be
\nu_h \approx 1\;,
\label{nu_final}
\ee
where the subscript $h$ refers to the direction of the random longitudinal field.

%%%%%%%%%%%%%%%%%%%%%%%%% Fig 5 %%%%%%%%%%
\begin{figure}[t!]
\includegraphics[width=1\columnwidth]{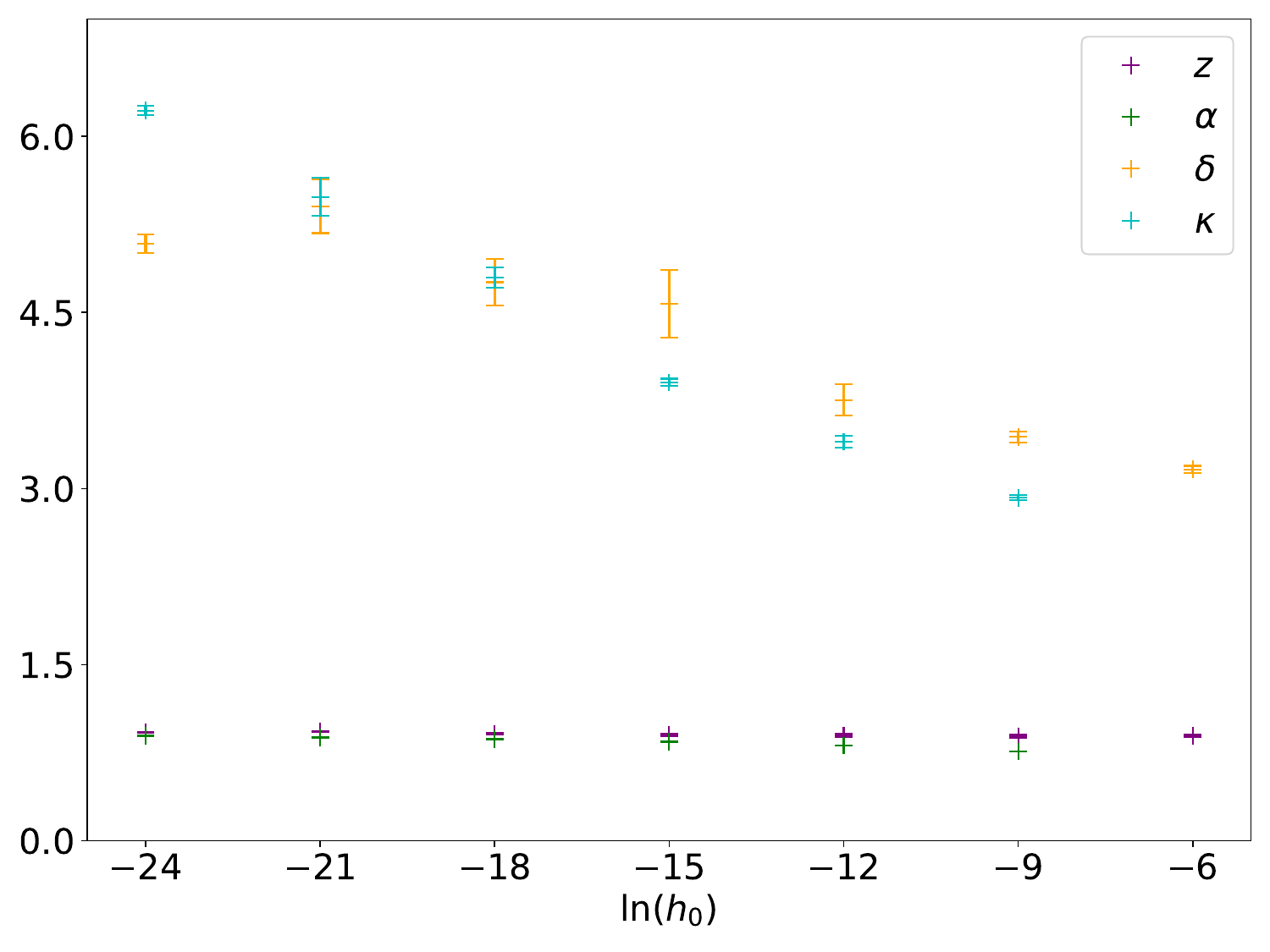}
\includegraphics[width=1\columnwidth]{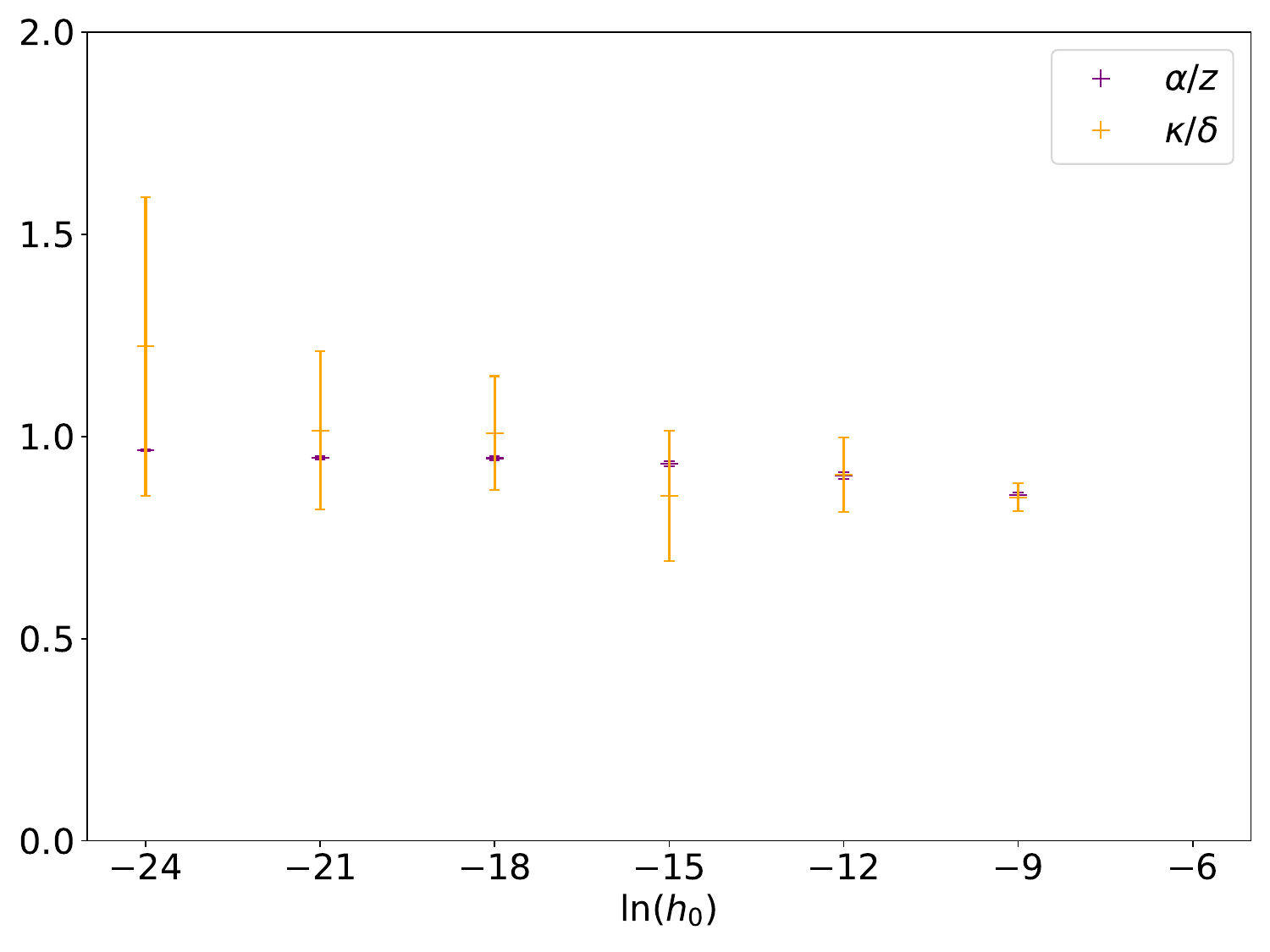}
	\vskip -0.4cm
\caption{Exponents along the special points, which define the relevant scaling direction. Upper panel: Estimated values of the local exponents in Eqs.(\ref{fix_L}) and (\ref{fix_h_0}) for different parameters of the distribution of the random longitudinal field, $h_0$. Lower panel: Ratio of the local exponents $\alpha/z$  and $\kappa/\delta$ for different values of $\ln h_0$. %These are calculated at the separation points.
}
\label{fig_10}	
\end{figure} 

\subsection{Behaviour of the log-gaps}
\label{sec:log_gaps}

We have also studied the distribution of the log-gaps, which is illustrated in Fig.\ref{fig_11} at $\ln h_0=-6$. For different sizes, the distributions are shifted (see the inset of Fig.\ref{fig_11}), and can be put to a master curve using the scaled variable $u=\epsilon L^z$. Here, the dynamical exponent corresponds to the value obtained from Fig.\ref{fig_8}. The master curve is well described by a Fr\'echet extreme-value distribution\cite{Galambos}:
%
%\be
%P_L(u;z)=\frac{1}{z}(u/u_0)^{1/z-1}\exp\left(-(u/u_0)^{1/z}\right)\;,
%\label{Frechet}
%\ee
%
%
\be
\ln P(\tilde{\gamma}-\gamma_0;z)=-\frac{1}{z}\tilde{\gamma} - \exp\left(-\tilde{\gamma}/z\right)+\ln(1/z)\;,
\label{Frechet}
\ee
with $\tilde{\gamma}=-\ln u+\gamma_0$ being the scaled log-gap variable and $\gamma_0$ is some constant, as shown in the main panel of Fig.\ref{fig_11}. For further discussions on the use of extreme-value statistics in the analysis of the gap-distributions in random quantum systems, see Refs.\cite{PhysRevB.73.224206,PhysRevResearch.3.033140}.

%%%%%%%%%%%%%%%%%%%%%%%%% Fig 3 %%%%%%%%%%
\begin{figure}[t!]
\includegraphics[width=1\columnwidth]{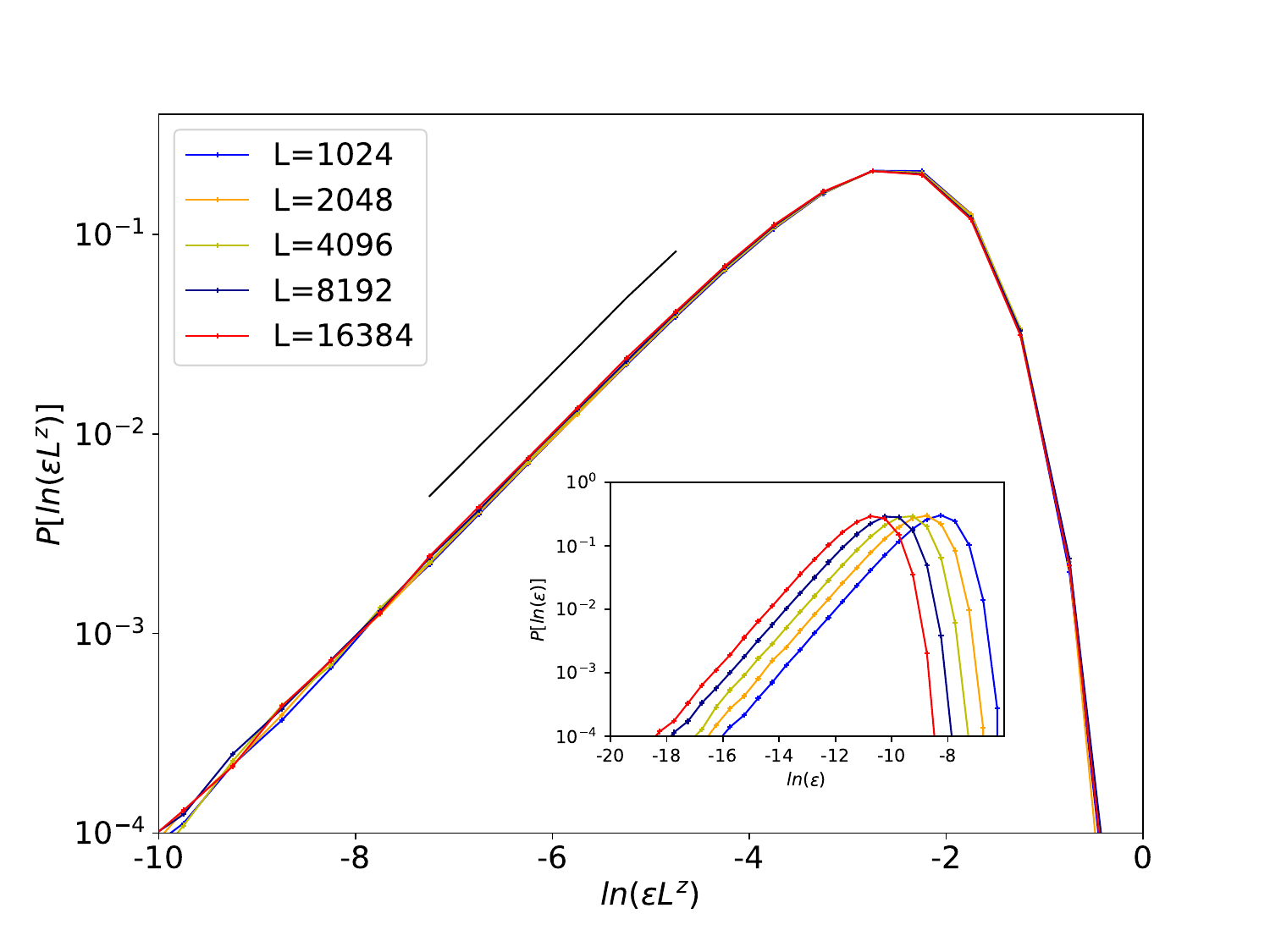}
	\vskip -0.5cm
\caption{Distribution of the log-gaps at $\ln h_0=-6$ for different sizes (inset), at $\Gamma_0=0.93$. The full line corresponds to the Fr\'echet distribution in Eq.(\ref{Frechet}) and the straight line has a slope $1/z$. Main panel: Scaled curves using the combination, $u=\epsilon L^z$, where the dynamical exponent is taken from Fig.\ref{fig_8}. }
\label{fig_11}	
\end{figure} 
%%%%%%%%%%%%%%%%%%%%%

The value of the dynamical exponent, $z$, depends on the distribution of the random longitudinal fields. The estimated values with the distribution in Eq.(\ref{eq:J_distrib}) having $\zeta=1$ are shown in Fig.\ref{fig_8}  for different values of the parameter $h_0$. According to this figure, $z$ appears to increase monotonously with decreasing value of $h_0$, having a saturation value of $z \approx 0.9$. Since $z<1$ the average susceptibility is not singular, but the non-linear susceptibility is a singular quantity. 

If we select a smaller value of the parameter $\zeta$, which measures the fraction of sites having random longitudinal fields \textcolor{black}{in Eq.(\ref{rho_i})}, it will result in a dynamical exponent $z>1$, as illustrated in Fig.\ref{fig_12} for  $h_0=1$ at $\Gamma_0^c=0.93$.

%%%%%%%%%%%%%%%%%%%%%%%%% Fig 5 %%%%%%%%%%
\begin{figure}[t!]
\includegraphics[width=1\columnwidth]{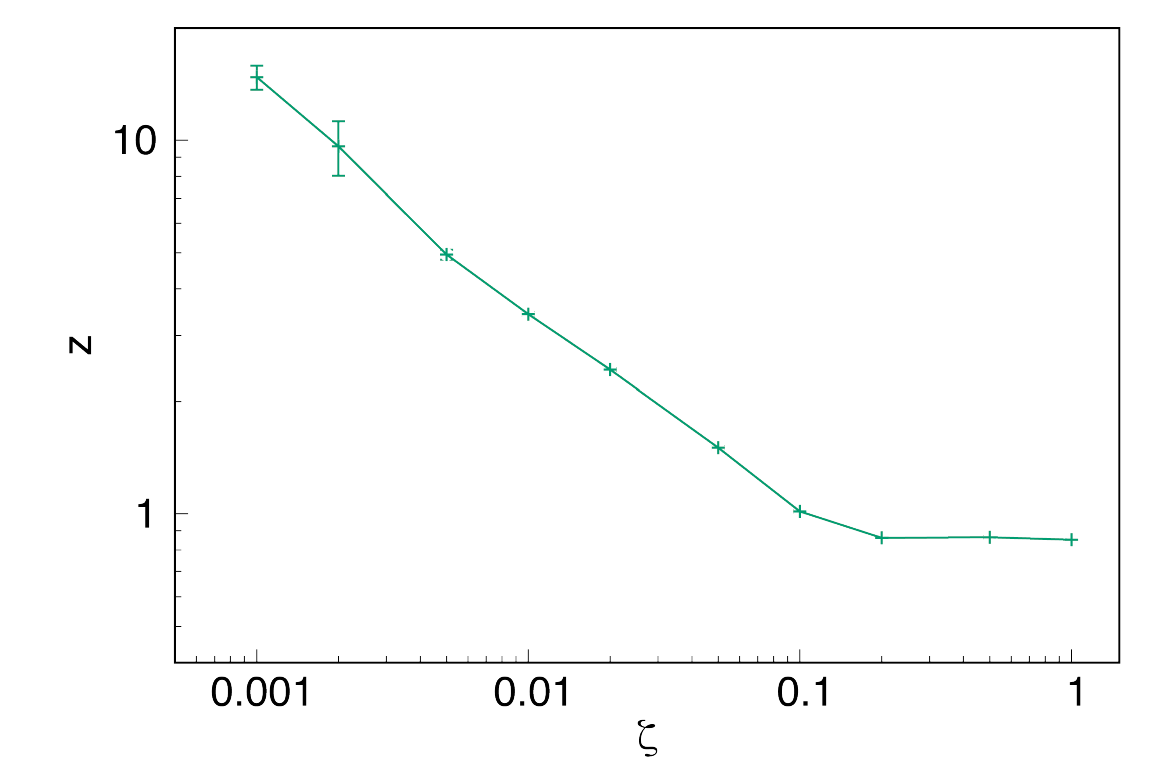}
	\vskip -0.4cm
\caption{The dynamical exponent, $z$, in such a random quantum Ising chain at $\Gamma_0^c=0.93$ in which with probability $\zeta$ there is a random longitudinal field with $h_0=1$, and $h_0=0$ otherwise, \textcolor{black}{see in Eq.(\ref{rho_i})}.
}
\label{fig_12}	
\end{figure} 
%%%%%%%%%%%%%%%%%%%%%

%\begin{table}[h]
%\begin{tabular}{|c|c|c|}
%\hline
% $p$& $z$\\
% \hline
% 1.& 0.64355823 \\
% 0.1& 1.1202892 \\
% 0.01& 3.61128469  \\
% 0.001& 9.08413083 \\
% \hline
%\end{tabular}
%\caption{The dynamical exponent $z$ in such a random quantum Ising chain in which there is an uniformly random longitudinal field with probability $p$.}
%\label{table_1}
%\end{table}

In this figure, the exponents have approximately a power-law dependence: $z(\zeta) \sim \zeta^{-\sigma}$, with $\sigma \approx 0.51(2)$. This result can be interpreted in the following way. In the first $n(\zeta) \sim 1/\zeta$ RG steps typically no random longitudinal fields are involved in the renormalization, while the typical strength of the log-couplings and log-transverse fields will be reduced by a factor of $f(\zeta) \sim n(\zeta)^{\psi}$, where $\psi$ is expected to approach $1/2$ for very large $n(\zeta)$, see in Eq.(\ref{psi}). This means that after the initial period of the renormalization the relative log-energy scale will be $|\ln \epsilon(\zeta)| \sim f(\zeta) |\ln \epsilon|$ and this relation is expected to hold until the last renormalization step. This way, the dynamical exponent following from Eq.(\ref{fix_h_0}) will be $z(\zeta) \approx z(1) f(\zeta)$ 
and $\sigma \approx \psi$.

\section{Discussion}
\label{sec:disc}
Understanding disordered quantum systems in the vicinity of their critical point is a challenging theoretical problem, since the collective behaviour is the result of quantum and disorder fluctuations in the presence of strong correlations. In a broad range of models, the critical behaviour is controlled by an infinite disorder fixed-point (IDFP) and the critical properties can be studied by the use of the strong disorder renormalization  approach. In the present paper, we considered a prototypical model, the random Ising chain in the presence of random longitudinal and transverse fields. Our study is motivated by the low-temperature properties of the compound ${\rm LiHo}_x{\rm Y}_{1-x}{\rm F}_4$, which is placed into a magnetic field which is transverse to the Ising axis. Using the SDRG method, we have studied the zero-temperature 
properties of the system. 

The critical behaviour of the system is governed by an IDFP, which is located at zero longitudinal field, $h_0=0$ and at $\Gamma_0=\Gamma_0^c$, using the random distributions in Eq.(\ref{eq:J_distrib}). Switching on the random longitudinal field, the ordered phase in the system disappears, and \textcolor{black}{in the renormalized system the couplings are very small while the transverse- and longitudinal fields are comparably very large. In such situation, the state of the system is trivial: it is a composition of (very weakly interacting) sites in random composite fields. Starting with a small $h_0$, the system will renormalize to one of these points only for larger transverse fields: $\Gamma_0>\Gamma_s$.  If, however, $\Gamma_0<\Gamma_s$, dominantly couplings are decimated, and when lengths are rescaled by a factor $\ell$, so that $\tilde{L}=L/\ell$, the typical renormalized transverse fields are $|\log \tilde{\Gamma}| \sim \ell$ and the longitudinal fields are
$\tilde{h} \sim h_0 \ell^{1/2}$. If at one step, the renormalized longitudinal field exceeds the value of the largest coupling, $\tilde{h} > \tilde{J}$, then the generated new term is typically larger than the actual energy-scale and at this point the original idea of the RG-process with continuously decreasing energy-scales will not be satisfied. At the last steps of the RG process, we have a system consisting of typical spin clusters  or domains of size $\ell \sim \min(L,h_0^{-2})$, which have log-couplings and log-transverse fields of typical  values $|\log \tilde{J}| \sim |\log \tilde{\Gamma}|\sim \min(L,h_0^{-2})$, while $\tilde h \sim O(1)$. For small $h_0$, we identify this region where the disorder is dominated by the classical random-field Ising model effects. The two regimes of the disordered phase noticed for small values of $h_0$ have significantly different characters and between those there is a cross-over region}, which starts at the IDFP. Below the cross-over the trajectories are attracted by fixed points which have classical random-field Ising character, whereas above the cross-over these scale to disordered quantum magnets. We have estimated the location of the cross-over region from the condition that at this point the value of the low-energy excitations is minimal. \textcolor{black}{We have shown that the position of the cross-over region is sharply defined in the limit of a small $h_0$.} We have estimated the correlation-length critical exponent along the special points which defines the relevant scaling direction and obtained a value $\nu_h \approx 1$, both for energy- and magnetization lengths. \textcolor{black}{Repeating the calculation along the line $\Gamma_0=\Gamma_0^c$ we obtained a different value for the critical exponent of the magnetization length, $\nu_{\mu} \approx 2$. We argue that this value agrees with the critical correlation length exponent  due to random transverse fluctuations and is connected to the fact that the point of reference has a distance from the relevant scaling curve which is proportional to $h_0$. }

We have also measured the value of the dynamical exponent, which is found to depend on $h_0$ and on the fraction of sites, $\zeta$, which are under the influence of the random longitudinal field, \textcolor{black}{see in Eq.(\ref{rho_i})}. In the case of $\zeta=1$, the dynamical exponent approaches a value $z \approx 0.9$, as $h_0 \to 0$. Since $z(h_0=0)$ is formally infinity, this means that the dynamical exponent has a discontinuity at $h_0=0$. We have shown that the dynamical exponent increases with decreasing value of $\zeta$, and it will diverge as $\zeta \to 0$, eventually leading to an IDFP. We have also shown that the distribution of the low-energy excitations are well described by the Fr\'echet extreme-value distribution.

Considering the model in higher dimensions, the RG phase-diagram in Fig.\ref{fig_1} remains unchanged in $d=2$, as there is no ordered phase in the classical random-field Ising model\cite{Binder1983}. On the contrary, in $d=3$, for small enough random longitudinal fields, there is an ordered phase\cite{PhysRevLett.35.1399,PhysRevLett.59.1829}  and the RG phase-diagram will have the expected form in Fig.\ref{fig_13}. Our aim in the future is to study in details the three-dimensional problem.

%%%%%%%%%%%%%%%%%%%%%%%%% Fig 1a %%%%%%%%%%
\begin{figure}[h!]
\includegraphics[width=1. \columnwidth]{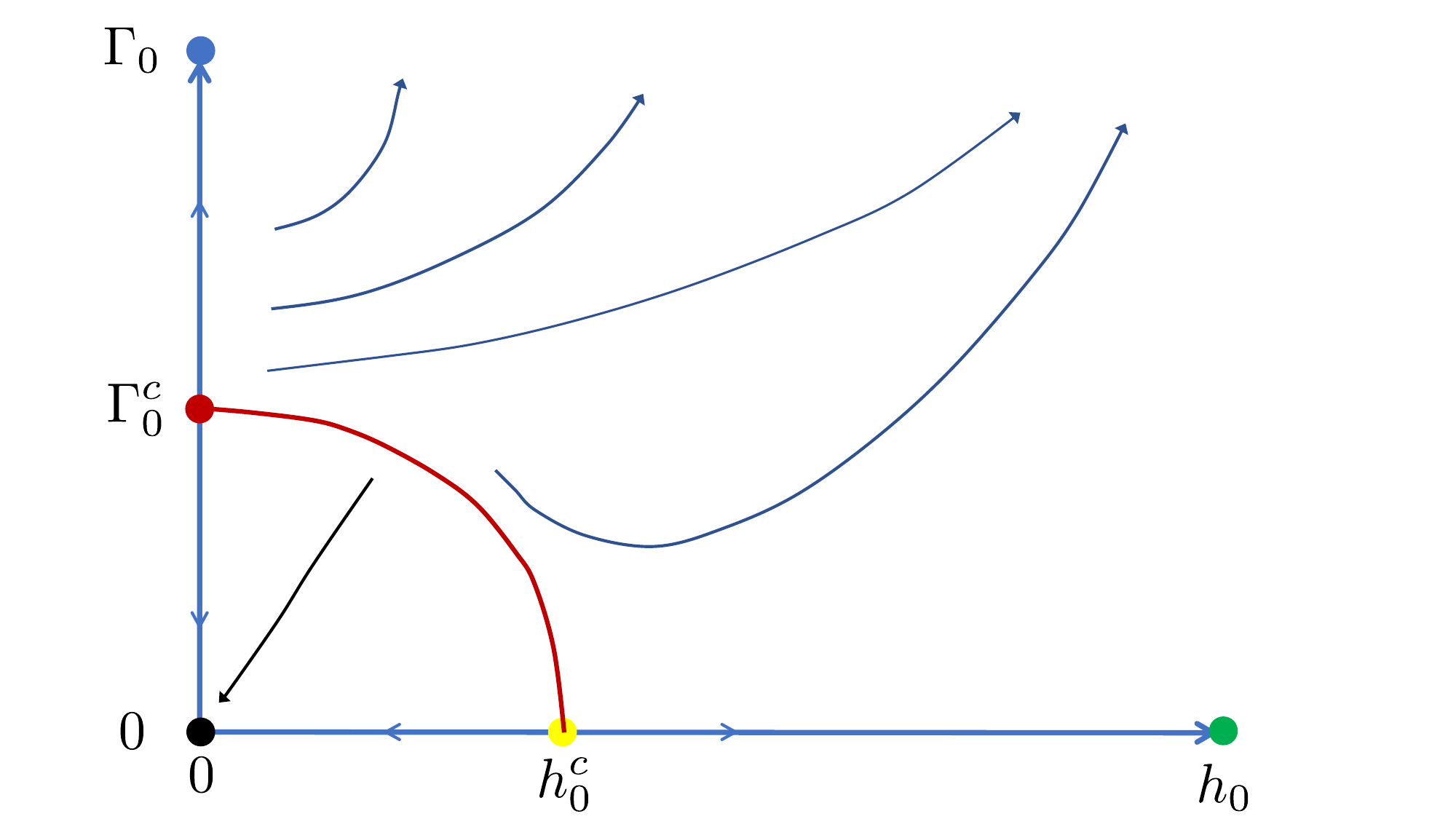}
	\vskip -0.3cm
\caption{Expected schematic RG phase-diagram for the three-dimensional model. At $h_0=0$ the IDFP (denoted by red circle) separates the quantum ordered phase ($\Gamma_0<\Gamma_0^c$) from the quantum disordered phase ($\Gamma_0>\Gamma_0^c$). At $\Gamma_0=0$ the ordered phase survives until $h_0<h_0^c$, denoted by a yellow circle. The  ordered and the disordered phases are separated by the red line. The RG-flows are illustrated by blue and black lines.}
\label{fig_13}	
\end{figure} 
%%%%%%%%%%%%%%%%%%%%%.

\begin{acknowledgments}
This work was supported by the National Research Fund under Grant No. K146736, and by the National Research, Development and Innovation Office of Hungary (NKFIH) within the Quantum Information National Laboratory of Hungary.
The work of IAK was supported by the National Science Foundation under Grant No.~PHY-2310706 of the QIS program in the Division of Physics. 

\end{acknowledgments}

\bibliography{Bibliography_quantrand}

\end{document}